# Measuring transnational social fields through binational link-tracing sampling

**Marian-Gabriel Hâncean**
Department of Sociology
University of Bucharest
gabriel.hancean@sas.unibuc.ro

**Miranda J. Lubbers**
Department of Social and Cultural Anthropology
Universitat Autònoma de Barcelona
mirandajessica.lubbers@uab.cat

**José Luis Molina**
Department of Social and Cultural Anthropology
Universitat Autònoma de Barcelona
JoseLuis.Molina@uab.cat

## Abstract

We advance binational link-tracing sampling design, an innovative data collection methodology for sampling from transnational social fields, *i.e.*, transnational networks embedding migrants and non-migrants. This paper shows the practical challenges of such a design, the representativeness of the samples and the qualities of the resulted networks. We performed 303 face-to-face structured interviews on sociodemographic variables, migration trajectories and personal networks of people living in a Romanian migration sending community (*Dâmbovița*) and in a migration receiving Spanish town (*Castellón*), simultaneously in both sites. Inter-connecting the personal networks, we built a multi-layered complex network structure embedding 4,855 nominated people, 5,477 directed ties (nominations) and 2,540 edges. Results indicate that the participants' unique identification is a particularly difficult challenge, the representativeness of the data is not optimal (homophily on observed attributes was detected in the nomination patterns), and the relational and attribute data allow to explore the social organization of the Romanian migrant enclave in *Castellón*, as well as its connectivity to other places. Furthermore, we provide methodological suggestions for improving link-tracing sampling from transnational networks of migration. Our research contributes to the emerging efforts of applying social network analysis to the study of international migration.

## 1 Introduction

Migration is not randomly distributed across the globe. Specific *binational migration corridors* can be identified, such as Mexico-US (the largest corridor between 1990 - 2000 and between 2000 - 2010) and Syria-Turkey (far the largest between 2010 - 2017; International Organization for Migration 2017). Inside such corridors, due to *migration networks* or *chain migration,* we can detect flows from one specific geographical area within a country of origin to a specific area within a destination country (Smith 2005). This network mechanism implies that once a small number of people from a specific area have settled in a certain destination area, it is easier for others to undertake the same trajectory. Migrants of former waves can pass information to help them start their migration project, find employment or housing, understand the national legislation and administrative frameworks.

Such regional migration corridors affect not only migrants, but also non-migrants and returnees, through exchanges of information, remittances, services, and "culture". To investigate these exchanges and their effects, scholars of transnationalism have proposed the term "*transnational social field*" (TSF), defined as "an unbounded terrain of interlocking egocentric networks that extends across the borders of two or more nation-states and that incorporates its participants in the day-to-day activities of social reproduction in these various locations" (Fouron and Schiller 2001: 544). TSFs capture "immigrants, persons born in the country of origin who never migrated, and persons born in the country of settlement of many different ethnic backgrounds" (Schiller and Fouron 1999: 344). Thus, TSFs are defined on the basis of migrants who move between geographically defined places of origin and destination, and only include return migrants and non-migrants insofar they are connected to the focal actors by a relevant social relationship (Lubbers, Verdery, and Molina 2018).

TSFs have fuzzy boundaries, both in terms of geography and membership. This is a challenge for constructing samples from such fields. First, migrant populations often lack a *sampling frame*, which means that the size and boundaries of the population (including the geographical dispersion) are unknown to the researchers (Heckathorn 1997). To this effect, they can be considered *hidden* or *hard-to-reach* populations (Heckathorn and Cameron 2017; Spreen 1992). Whereas in the case of *known populations* (*i.e.,* for which a sampling framework exists), traditional probability sampling methods can be applied, these methods lack efficiency in producing reliable samples for *migrant populations.* Additionally, efforts to quantitatively describe migrant populations are severely restricted by a wide range of other challenges, including residential mobility, low availability for home interviewing, reluctance to research participation and to revealing personal data, lack of trust, high sensitivity to specific research topics, official language barriers, cultural differences, legal status, social security affiliation (Font and Méndez 2013). On top of that, TSFs do not only focus on migrant populations but also on non-migrants and returnees insofar they are connected to the focal migrants. As it is a priori unknown to researchers who is and who is not associated with migrants in the specific area of destination, these individuals can only be indirectly sampled, through the referral of others.

Solutions to the impracticability of traditional probability designs to the study of migrant populations are still in an infant stage of development. At the same time, despite the network character exhibited by the migration processes, substantive research into these networked processes have proven to be rare until recently (Bilecen, Gamper, and Lubbers 2018). In this context, our study contributes to the recent efforts of quantitatively describing migrant populations and migration networks (Lubbers et al. 2018) by adopting a network research oriented design (*i.e.,* employing a chain-referral data collection strategy). The design intends to sample from the TSFs and to increase the understanding of migration processes and patterns. Our research extends already existing methodologies (Merli et al. 2016; Mouw et al. 2014; Mouw and Verdery 2012) in terms of data collection process (*i.e.,* implementation of simultaneously and not sequentially multi-sited data collection) and of providing more thorough description of migrants' networks (*i.e.,* elicitation of more network contacts, relationships among contacts, collection of diverse attribute data as to increase the accuracy and robustness of identifying across nominations the unique individuals within the network). Thus, we not only provide replication results (with reference to previous research work), but also bring forth new insights on migration within transnational networks.

To advance migration research, this paper aims to move beyond the currently employed research frameworks by deploying an innovative quantitatively oriented methodology, that, unlike previous research, is focused simultaneously on origin and destination places. As an application, we studied the TSF created by Romanian immigrants from *Dâmbovița* (a Romanian county with a population of nearly 530,000 people, situated at 78 km North-West of Bucharest, Romania) to *Castellón* (a Spanish province of nearly 577,000 inhabitants, situated on the Mediterranean coast, where 11% is Romanian). The method is based on network sampling, a sampling strategy that uses social networks to obtain convenience and representative samples from hidden populations. However, so far, descriptions of the practical implementation of such methodologies are lacking, despite their relevance for guiding future studies (*e.g.*, controlling the homophily in the nomination patterns, solving for the unique identification of participants, assessing the representativeness of the resulting samples). Therefore, this paper describes the implementation of the methodology for the empirical study of TSFs and the characteristics of the resulting sample. Additionally, while sampling is the primary objective of network sampling methodologies, in the context of TSFs, an equally important objective is the detailed study of the functioning of migration networks. Therefore, we are also interested in the type of network that the binational link-tracing design reveals. Our research questions are threefold: (1) What are the *practical challenges* of implementing the binational link-tracing design empirically? (2) How *representative* is the sample obtained with the binational link-tracing design? (3) What are *the qualities of the migration network* visualized with the binational link-tracing design?

The structure of the paper is as follows. Firstly, we review the literature on network-oriented sampling methods for hidden populations. Secondly, we briefly introduce network approaches to measure TSFs. Afterwards, we describe our research design (a binational link-tracing variant) implemented for measuring the TSF wherein Romanian migrants in or returned from Spain (Castellón) as well as their social contacts (relatives, friends and acquaintances) are embedded. We then present the results obtained after sampling from the hidden population of Romanian migrants (the major demographic characteristics of the study participants, the structural and compositional features of the measured TSF). Finally, the paper discusses the theoretical and practical implications of our study, some limitations, and future directions.

## 2 Sampling from transnational social fields

### 2.1 Network-oriented sampling methods

Migrants and non-migrants in TSFs are typically considered a hidden or hard-to-reach population, *i.e.*, a population for which the degree of access for collecting data is low. Due to the impracticability of non-network probability sampling designs to the study of hidden populations, other methods have been deployed. We refer here to *network-oriented sampling methods* (or *link-tracing sampling* methods), which essentially implement chain-referral strategies for collecting data. Initially, link-tracing sampling designs, such as snowball methods (Goodman 1961), were used to construct networks and study social structures (Coleman 1958; Heckathorn 1997, 2002; Heckathorn and Cameron 2017; Spreen 1992). Network oriented sampling designs were rapidly transferred to the study of hidden populations due to their capacity of locating affiliated members (Heckathorn and Cameron 2017). Specifically, as Spreen (1992) argues, network-oriented sampling methods use a link-tracing or chain-referral strategy of collecting data (*i.e., chain data*) and allow for eliciting members in hidden populations (such as population of migrants). Researchers' appeal to this specific class of sampling methods could be explained by the effectiveness of *locating members* of hidden populations, as well as by the superiority in rapidly *increasing* the number of members of a target population in a sample. The most popular non-probability form of link-tracing method is *snowball sampling* (Heckathorn 2011). This method was described by Goodman (1961) as implying $s$ stages and $k$ names. Precisely, a small, randomly selected set of individuals from a given population is used as the first phase of the sampling procedure (the *seeds*). Next, each individual in the set is asked to name $k$ individuals in the population who are not in the randomly selected set. The $k$ people form the second phase and are asked to further name $k$ individuals. The $k$ people who are not in the first and second phase are then asked to name $k$ different individuals. The procedure continues until $s$ stages or a specific sample size is achieved (Heckathorn 2011) .

Despite Goodman's (1961) description, the term snowball sampling is currently used for any method that starts with a small number of (usually not randomly selected) seeds, and asks them for referrals (as many as they can give) until the desired sample size is obtained or saturation is reached. Thus, the number of names per interviewee, the number of stages, and the precise referral chains are not controlled. In effect, despite providing a higher degree of coverage for the cases of hard-to-reach populations (compared to traditional probability sampling methods), snowball designs typically produce convenience samples. Contrary to initial claims that snowball sampling can be used to make statistical inferences (Goodman 1961), multiple sources of biases were shown (Erickson 1979; Heckathorn 2002, 2011). Precisely, firstly, it was argued that the *initial sample is unlikely to be representative*. Among others, the number of seeds is often too small, and their participation often involved *volunteering*. Second, chain-referral samples were suggested to be *biased toward more cooperative participants*. Thirdly, it was suggested that the attributes of seeds impacted upon additional participants through *homophily*, which is especially troubling when initial subjects are not randomly selected. Fourthly, participants tended to protect their friends by not referring them, particularly when privacy issues are involved, *i.e.*, a tendency called *masking*. Fifthly, as referrals occur through network ties, individuals with larger personal networks have greater chances of being selected thus being oversampled. Because of these biases, snowball and similar chain-referral samples have been appraised as *convenience samples* (*i.e.,* non-representative samples).

## 2.2 Respondent driven sampling methods

The efforts of transforming link-tracing / chain-referral designs into probability sampling methods are manifest in the work on *respondent driven sampling* (RDS). RDS illustrates a class of methods aiming to convert chain-referral sampling into a method of good estimability, by reducing some of its critical biases (Heckathorn 1997, 2002, 2011; Heckathorn and Cameron 2017). RDS is shown to be based on Markov chains as well as on a dual system of incentives to drag behavioral compliance on the part of subjects from the target population. According to Heckathorn, firstly, by implementing a Markov modeling peer recruitment process (*memoryless recruitment*), as sample increases one wave after another, an equilibrium sample composition is rapidly achieved. That means seeds' biases are eliminated. Secondly, RDS typically employs a dual incentive system: rewards for being interviewed – *primary incentives* –, as well as for recruiting others – *secondary incentives* (the latter rewards are effective for recruiting less cooperative subjects). Thirdly, an RDS sample is reported to be unbiased whether the homophily of each group is equal or whether the network size of the participants is controlled. Fourthly, study participants are not required to identify their peers but to recruit them. In effect, *the masking* bias is said to be reduced as respondents are given the liberty to allow peers to decide for themselves whether they participate to the study. Fifthly, *recruitment quotas* (*i.e.,* the fixed maximum number of names respondents are asked to recruit) have been shown to be effective means for reducing the impact of subjects with large personal networks on the recruitment patterns.

By convention (Heckathorn 1997), RDS starts from a set of seeds that are financially incentivized to recruit peers. The same system of incentives is applied to all recruits, irrespective of their status - *seeds* or *referrals*. The chain-referral mechanism should work only with objective verifiable criteria for assessing membership in the targeted population. Very clear traits for establishing membership are useful for cases of subject duplication (*i.e.,* multiple participation under different identities) or of subject impersonation (*i.e.,* cases when a subject pretends to be one of her peers just to collect the incentives). Generally, sampling is completed when either the targeted population is saturated, or a specific size and content of the sample has been reached. Evidently, RDS can be practiced only for populations which exhibit a contact (relational) pattern; there should be ties connecting peers. Furthermore, it is only possible for those cases wherein a trait defining membership in the population is available for objective verification.

In addition to the sampling procedure, (Heckathorn 2002) advanced an RDS population estimator that accounts for both the differences in homophily across groups and the variation in the size of the personal networks (*i.e.,* subjects' number of social contacts). The development of this estimator was critical, as the organization of social (network) structures is generally *homophilous* (McPherson, Smith-Lovin, and Cook 2001), *i.e.,* individuals tend to interact with similar others. Consequently, in practice, it was observed that homophily exponentially inflates the standard errors. That was solved by subdividing samples in homophily breakpoints to control for the variability of the estimates.

Heckathorn's RDS estimator is asymptotically unbiased (*i.e.,* biases are only of the order of 1/n, where *n* designates the sample size) under the following assumptions (Salganik and Heckathorn 2004): *i*) each subject is connected by at least one link to the rest of the targeted population (*network embeddedness*); *ii*) all members of the targeted population belong to a single *component, i.e.,* every member of the targeted population is part of one global network; *iii*) sampling is performed with replacement (sampling fraction is as small as possible); *iv*) the personal network size is accurately reported by each respondent; *v*) each subject randomly recruits from her network (satisfying this assumption, respondents are inversely weighted by the size of their personal network); *vi*) each subject recruits a fixed number of peers.

Another way of approaching link-tracing designs involves adaptive sampling, *i.e.,* information collected during the sampling process orients future sampling work (Thompson and Collins 2002). Specifically, only respondents who satisfy specific criteria are asked to recruit peers. Estimators from adaptive sampling are valid as long chain-referral waves reach saturation and seeds are randomly selected. This method, which implies maximum likelihood estimation, is limited to instances wherein initial respondents can be randomly drawn and exhaustive link-tracing is feasible in the population.

Other estimators have been developed using egocentric data collected via RDS, *i.e.,* each respondent, who is connected to her recruiter, provides information on the composition of her personal network or about the proportion of her peers sharing specific attributes (Lu 2013). Particularly, this method estimates not only the inclusion probability for every respondent but also for any of her alters or peers. The transition probabilities (*see* Heckathorn 1997, for a discussion) are computed based on each respondent's network composition (alters and their attributes). Using simulations, it was shown that the ego network approach provides estimates for which two important biases were controlled, *i.e.,* differential recruitment (different patterns of recruitment) and peer underreporting. The main limitation assigned to this method refers to the respondent being able to accurately provide information about the number of alters, which in practice is highly questionable.

## 2.3 Link-tracing sampling from transnational social fields

Currently, there is a wide consensus among scholars that both migrants and non-migrants' lives are, to variant degrees, transnational (Faist, Fauser, and Kivisto 2011). Cross-border activities and transnational practices, such as communication (via telephones, Skype, WhatsApp, or social media platforms), travel, flow of money and other forms of remittances (Levitt 2001) have been used as indicators of the intensity of transnationality in individuals' life (Mouw et al. 2014; Verdery et al. 2018). As previously pointed out, migrants live multi-sited lives that include not only their home and destination places but other sites worldwide (Levitt and Jaworsky 2007). This way of living connects migrants to other migrants and non-migrants, and, in effect, produces "multiple interlocking networks of social relationships" (Levitt and Schiller 2006: 1009) or *transnational social fields i.e.,* "networks of networks that stretch across border-states" (Schiller 2005: 442). Inside these social structures, lives of non-migrants are also transformed despite their immobility (Levitt 2001).

The research on TSFs traditionally tended to disregard the potential benefits of incorporating social network analysis into its methodological apparatus. However, recently, efforts have emerged to represent TSFs through the use of social network analysis tools (Bilecen et al. 2018). (Lubbers et al. 2018) identify four classes of approaches based on the unit of analysis: *the personal network approach* (focused on individuals), *the household survey approach* (focused on households), *the simultaneous matched samples methodology* (focused on dyads), and *the binational link-tracing design* (that encloses a community focus). While the first approach enquires about network members regardless of where they live, it does not sample these network members for further investigation. The other methods, in contrast, tend to invite one or multiple network members of respondents to participate in the research to investigate for example the effect of migration experience of relatives on migration intentions, the transnational exchange of remittances and services, or the configurations of care relationships in transnational families.

The binational link-tracing design (Merli et al. 2016; Mouw et al. 2014; Mouw and Verdery 2012) is heavily built on the simultaneously on-going methodological efforts of transforming chain-referral designs into probability sampling methods (Heckathorn and Cameron 2017; Mouw and Verdery 2012; Verdery et al. 2015, 2017). In a nutshell, the binational link-tracing design deploys RDS by sampling individuals both in the sending and receiving places of a migration corridor. Specifically, in the first phase, the elicitation of the TSF starts with a small convenience sample of seeds in the area of destination, after performing ethnographic fieldwork in the community. Individuals in the initial sample nominate other people in the origin and destination places. On the one hand, they are asked to describe their personal network, by eliciting a list of network members (friends, family and acquaintances in the area of origin, the area of destination, returned migrants) and enquiring about their characteristics. Respondents are not asked whether the network members are connected among each other; if the sampling fraction is high enough, some of this information should be available through the link tracing network. On the other hand, respondents are asked to give a small number of names of people in both the area of origin and destination who might want to participate in the survey (referrals). The referrals in the destination place are then asked to participate in the survey. This procedure is continued until the desired sample size is reached in the place of destination.

In the second phase, data are collected in the community of origin, based on the referrals of the participants in the destination area. Again, information on their personal networks is elicited. Mouw and Verdery (2012) applied this technique to study a migrant community spanning three regions: The Raleigh-Durham-Chapel Hill area of North Carolina; Houston, Texas; and Guanajuato, Mexico, with more than 600 respondents in total.

After data collection, the information is combined to construct a network embedding all the interviewees and their referrals. To do so, it is essential that all individuals are uniquely identified. Therefore, respondents were asked to give the first four letters of the first names and of the surnames of themselves and their network members, without affecting respondent compliance due to privacy concerns. The authors have later also successfully experimented with other identification techniques, namely by using the last four digits of nominees' phone numbers (Merli et al. 2016). Once data belonging to unique individuals are matched, the identifiers can be substituted for others for complete anonymization. Among others, the authors showed that the network underlying the TSF is an important vehicle for opinion formation about migration, and how transnational communication is affected by both individual and network characteristics.

As Lubbers et al. (2018) stress, the resulting network is only a sample or a part of the total TSF; inferences about the whole TSF could subsequently be derived through statistical or mathematical modelling (Verdery et al. 2017). Moreover, as Mouw and Verdery (2012) only asked about people living in the communities of origin and destination, individual transnationality could not be estimated in general, but only with regard to the given corridors. In our study, we grasp the approach introduced by Mouw and Verdery (2012), but adapt the methodology. The following section describes this methodology, while emphasizing its distinctive features and commonalities in relation to previous endeavors (*i.e.,* the work of Mouw and colleagues).

## 3 The Implementation of the binational link-tracing design

This section introduces and describes our research design. It was implemented for measuring the TSF created by Romanian migrants *in* or *returned* from a bounded area in Spain (*Castellón*), that also included their social contacts (relatives, friends and acquaintances) in Spain, Romania and elsewhere. We provide details on the studied population, the sampling procedure and the reward system for participation, on the questionnaire, and the unique identification of individuals. We also explain how various biases (initial sample biases, masking biases, social desirability) were addressed. Additionally, we shed light on the specificity of the current methodology as well as on the commonalities with previous work.

### 3.1 Romanian migration to Spain

Migration inside the European Union (EU) is highly dynamic (United Nations 2017). Currently, approximately 20 million European citizens live in a EU country in which they were not born. In 2017, nearly 4.0% of the EU citizens of working age (20-64) were residing in another EU member state – a share which increased from 2.5% in 2007 (Eurostat 2018). Romania is among the *top 20* countries in the world with the largest diaspora populations (United Nations 2016). Among the EU citizens of working age, Romanians have been the most mobile, –with 19.7% of the population living in another EU member state (Eurostat 2018). As of January 2018, Romanians were estimated to be among the top five most numerous foreign populations in, for instance: Italy (23% of the total foreign population), Spain (15%), Hungary (14%), Slovakia (9%), Portugal (7%) (Eurostat 2019). Since 2000, Romanian migration trajectories with the largest annual increase have been directed towards Italy and Spain (United Nations 2017). Consequently, it is no surprise that, on January 2018, the Spanish Institute of Statistics registered more than 675,000 Romanian residents in Spain, *i.e.,* the largest EU foreign population and the second largest foreign population after the Moroccans. In parallel, the Italian National Institute of Statistics reported a tally of more than 1,190,000 Romanian residents in Italy, the largest foreign population.

As already reported (Molina et al. 2018), Romanians in Spain are geographically unevenly distributed, being concentrated in geographically bounded areas (*i.e.,* migrant enclaves). One of these Romanian migrant enclaves is established in *Castellón* and accounts for at least 11% of the total population in this region. Romanians who had firstly arrived in *Castellón* were predominantly from *Dâmboviţa* (Bernat and Viruela 2011), a Romanian county with a population of 498,826 people (the population of 18 years and older has a size of 406,598; 49% male, 51% female; as of July 2017, Romanian National Institute of Statistics) situated at 78 km North-West of Bucharest, Romania. The steep increase of the migration flux of Romanians to Spain has been underpinned, since 2000, by *institutional factors* (*e.g.,* recurrent processes of regularization in Spain, free mobility due to Romania adhering to EU since 2007, Spanish immigration policies, aging of Spanish population etc.) and *linguistic proximity*. As a result, many Romanians from *Dâmboviţa* chose Spain as their destination place. As of January 2018, the number of inhabitants of the province of Castellón with Romanian nationality was 38,231, and the number of inhabitants of 18 years and older 30,880 (among the 18+, 47% males, 53% female; average age 40.6 years, $SD$ = 12.2; (Instituto Nacional de Estadistica 2017)). Romanians rapidly became the nationality with the largest number of residents and with the highest number of employees with a formal contract in this area (Bernat and Viruela 2011).

## 3.2 Sampling and procedures

Between November 2017 and July 2018, we conducted face-to-face structured questionnaire-based *pen-and-paper personal interviews* with 303 participants in two sites: 149 in *Castellón,* and 154 in *Dâmboviţa*. Three classes of respondents were sampled: *migrants in Spain* (Romanians living in *Castellón,* Spain), *return migrants* (Romanians who previously had lived in *Castellón,* Spain, but returned to *Dâmboviţa,* Romania) and *non-migrants* (people living in *Dâmboviţa,* Romania, who never migrated to Spain). The interviews were conducted, in parallel, by international researchers affiliated to the study. In *Dâmboviţa* (Romania), the fieldwork was undertaken by a team of five scholars who conducted all interviews in Romanian. A team of three scholars of different nationalities undertook the fieldwork in *Castellón* (Spain), conducting 89 interviews in Romanian and 60 in Spanish. Participants were allowed to freely choose the physical places of their interviews to make them feel comfortable (*e.g.,* home, pubs, restaurants, public gardens, on the street). Interviews were scheduled using the means suggested in advance by the participants, such as: telephone, WhatsApp or Facebook. After each interview, brief reports were written up by the researchers. The coordination of data collection process as well as solving for administrative tasks were carried out through both communication technologies (*e.g.,* Skype meetings, WhatsApp, Email and Voice Calls) and on-site face-to-face meetings. Figure 1 illustrates the time evolution (in months) of the number of conducted interviews.

We deployed a *link-tracing* sampling design, a procedure essentially based on a chain-referral way of collecting data. Prior to sampling, a year of ethnographic fieldwork research had been conducted in *Castellón,* which provided important insights on the Romanian migrant population, such as its structure and social organization, its geographical mapping etc. As previously argued (Watters and Biernacki 1989), the quality of the sampling from a hidden population is heavily affected by the accuracy and the comprehensiveness of the ethnographic mapping. Consequently, in our case, building on the field reports of the ethnographic research stage, a number of nine *seeds* (the sample of initial subjects, *i.e.,* Romanian migrants) living in *Castellón* was *purposively* selected (Saunders, Lewis, and Thornhill 2012).

The seeds were selected to be as *heterogeneous* as possible to ensure that: a) the TSF surrounding the Romanian migrant community in *Castellón* was widely and extensively explored; b) the chain-referral network linkages did not collapse into a single network component after only a few waves. The heterogeneity of the initial number of seeds was achieved by using relevant and critical demographic features such as *sex, marital status, age, parenthood, level of education, religion,* and *work status.* These features were indicated as essential for the link-tracing sampling process by the insights of the ethnographic fieldwork undertaken in *Castellón.* Additionally, the seeds were selected from sub-groups within the Romanian community with little connection between them.

Each of the nine respondents and the subsequent referrals, after being interviewed, was asked to provide contact details on *three persons* (relatives, friends, and acquaintances) living in *Castellón* and on *three persons* living in *Dâmboviţa.* Respondents were informed to nominate as referrals only at least 18-year-old people with Romanian nationality. Eligibility for participants recruited in *Castellón* included residence for at least six months. The data collection process proceeded in a chain-referral way or through *referee – referral* network linkages: from the *seeds* to the *first wave* of respondents, then from the first to the *second wave*, then from the second to the *third wave* etc. until a target sample size was attained.

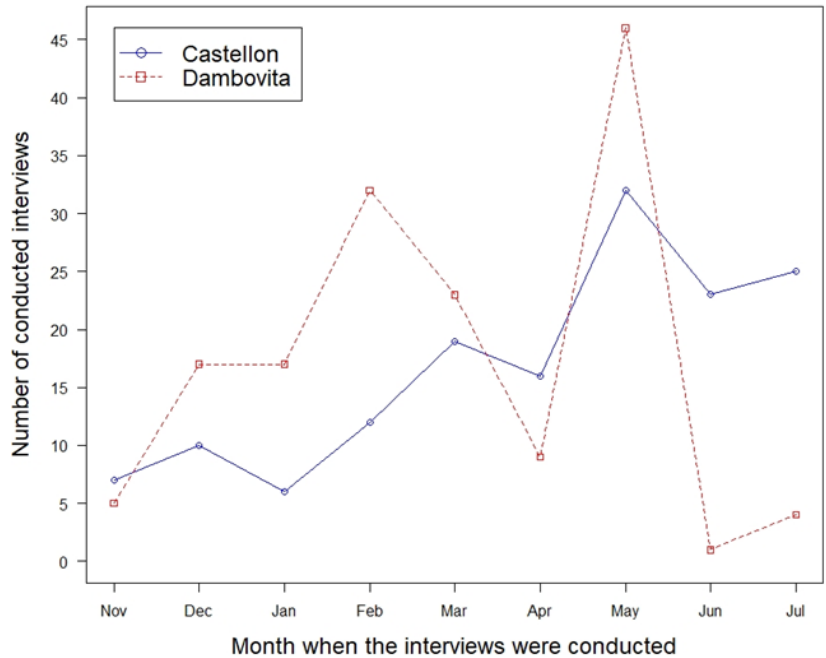

**Figure 1:** The process of link-tracing data collection, by month and location. *Note:* The lines are indicating the time variation of the conducted interviews (red line for the *Dâmboviţa* site, while blue for the *Castellón* site).

Participants to the study received monetary incentives both for accepting to be interviewed and for helping the research team in recruiting other people. Specifically, respondents (referees) were not required to just *identify* their peers but asked to *recruit* them *per se* into the research. In this way, a consistent reduction of *the masking bias* was expected, as explained before. Members of the research team were allowed to contact the referrals only after receiving the confirmation from the referees. Due to the inexistence of similar previous link-tracing sampling studies in the context of Romanian migration, participation and recruitment rewards were set at ten Euros, in a *face validity* fashion. It was estimated that the amount of money was, on average, sufficiently high to induce participation, and at the same time not too consistent to generate *social desirability effects, ethical problems of coerciveness,* or enlisting persons not part of the hidden population. Falsely claiming membership of the study population was controlled and validated by screening both the responses provided to the questionnaire items and the information collected from other participants.

We collected data in the two sites *simultaneously* as we wanted the time between referral and interview to be small to avoid respondents falling out. Identification data on the referrals (peers recruited by already interviewed participants or *referees*) was exchanged by the two research teams in a *ping-pong* game manner. Information, provided by the referees living in *Castellón,* on referrals living in *Dâmboviţa,* was electronically transmitted by the *Castellón team* to the *Dâmboviţa team,* and vice-versa. This electronic transfer of personal contact data was governed by a pre-defined protocol for data anonymization (as a part of a general study Ethics protocol approved by the ethical review board of the Autonomous University of Barcelona). The identification information (full name and contact details, such as phone number, Facebook account or email) was, afterwards, encoded using an alpha-numeric system: the first three letters of the name, the first three letters of the surname and the last four digits of the phone number. Before being interviewed, each participant was informed about the research, asked for his / her content to participate in the research, and (if consenting) signed a written consent form. If consent was not given, referrals were not interviewed.

We theoretically set a general sample size target of at least 300 interviews, approximately evenly split by the two sites: *Castellón* and *Dâmbovița*. We started with fewer seeds and extended the number when we found that six useful referrals were hard to get. Eventually, we needed nine seeds to accomplish a volume of 303 valid interviews (149 interviewees in Spain and 154 in Romania). The impact of the purposively selected sample of seeds on subsequently selected subjects, *i.e.,* the biases of the non-randomly selected initial seeds, was shown in the literature to be filtered out due to the attained long chains of referee - referrals. In our case, the resulting link-tracing network (*i.e.,* the network obtained after interconnecting the participants through the link-tracing) has a maximum wave length of 16, varying from one to 16 between seeds.

### 3.3 The questionnaire

We designed and separately applied three customized questionnaires for *migrants in Spain* (people living in *Castellón,* Spain), *return migrants* (people who previously had lived in *Castellón,* Spain, but returned to *Dâmbovița,* Romania) and *non-migrants* (people living in *Dâmbovița,* Romania, and who never migrated to Spain). Despite their customization, the questionnaires included a core-set of items for all study participants to allowed for comparisons across the three groups of subjects.

The questionnaires were devised in English and translated afterwards into Romanian and Spanish. To control the accuracy, validity and the quality of the translation, *forward* and *backward* translations were employed (Guillemin, Bombardier, and Beaton 1993). The pre-final versions of the questionnaires were assessed in an *expert committee* fashion (Beaton et al. 2000) and *pilot-tested* (Perneger et al. 2015) on a sample of five participants. After subsequent revisions, actions for ensuring the validity and reliability of the final version of the questionnaires were taken (Crocker and Algina 2008).

The questionnaires had several *blocks* of items. The first block registered participant's identification data, such as *participant's alias* (to ensure anonymization, participants' identity was encoded using an alpha-numeric system: first three letters from name, first three letters from the surname and the last four digits of the telephone number), *place of residence, sex, and date of the interview*. The second, third and fourth blocks enquired respectively about respondents' attributes (*birth year, marital and parenthood status, level of formal education, work status and place,* and *religion), life in Romania* and *migration experience to Spain* (e.g., *work experience, decision on migration, mobility and migration experience, properties owned in Romania and other countries (Spain, included), circulation of remittances, cultural consumption, social identity perceptions, satisfaction with life*) and *institutions (organizations) currently supporting respondents' migration experience* (if applicable).

In the fifth block, respondents were asked to elicit a specific number of personal contacts (relatives, friends and acquaintances); the so-call *name generators* allowing the construction of a personal network for each participant. Precisely, based on their place of residence (either *Castellón* or *Dâmbovița*), respondents were asked to elicit: a) maximum ten friends and acquaintances living in the current place of residence (*Castellón* for migrants and *Dâmbovița* for non-migrants and returnees); b) maximum five relatives living in *the current place of residence*; c) maximum five relatives, friends and acquaintances who had lived in *Castellón* but now live in Romania; d) maximum five relatives and five friends and acquaintances who live in the other place of the TSF (*Dâmbovița* for migrants and *Castellón* for non-migrants and returnees); e) maximum five relatives and five friends and acquaintances living in other places than *Castellón* and *Dâmbovița.* The application of the five name generators could theoretically elicit a maximum of 40 network members ("alters" in personal network research): 15-20 relatives and 20-25 friends and acquaintances. We did not collect the full names of network members, but only the first three letters of their first name, the first three letters of their last name, and the last four digits of their phone number, in correspondence with respondents' identifiers. Additional questions were used to collect information on the elicited alters (these questions are called "name interpreters" (McCarty et al. 2019), such as their attributes: *sex, occupation* and *religion,* and the respondent's relationship with him/her, such as: *the nature of the relationship* (*e.g.,* workmates), *duration* (in years), *emotional closeness*, and *communication frequency*. In the last section of the network module, we measured relationships among network members. In line with the suggestions available in the literature for reducing *respondent burden* in such questions (Golinelli et al. 2010; McCarty et al. 2007; Merluzzi and Burt 2013), we randomly sampled nine alters from those originally elicited to measure network structure. Respondents were asked to mention, for each pair of sampled alters, whether they *knew each other and could contact each other independently of the respondent*.

In the last block of the questionnaire, respondents were asked to recruit referrals eligible to participate to the research, *i.e.*, to ask them whether they would be willing to participate in the research, either on the spot or after the interview. In some instances, the recruited people had been also nominated as a result of the application of the name generators from *block 5*.

## 3.4 Interconnecting personal networks and the structure of the data

To build a single, *multi-layered network* or, in other words, the *network of networks* (or *a representation of the TSF*; the interconnecting of the link-tracing participants as well as their nominees, *i.e.,* referrals and alters), we uniquely identified each individual in the research, taking into account that an individual could appear multiple times in the data (*e.g.*, as a network alter of one respondent, a referral of another respondent and ultimately as respondent). The identification implied an extremely cumbersome and tedious procedure. Firstly, all nodes received an alphanumerical code in the data collection process. As indicated before, alphanumerical coding was derived from the first three letters of the name, first three letters of the surname and from the last four digits of the phone number. In some cases, due to missing data (*e.g.,* the alters for whom egos were not able to provide a phone number, or a last name), special coding was assigned. Secondly, the allocated alphanumerical codes were subject to a data cleaning and validation process. The process was meant to detect and correct corrupt or inaccurate records either due to data entry errors or conflicting coding (*e.g.,* in some cases, two different nodes were assigned the same code, whereas in others, the same node was allocated different alphanumerical codes). The data cleaning process was conducted using various methods: from manual screening and the use of the *RecordLinkage R package* (Sariyar and Borg 2010) in the initial stages, to employing Microsoft *Excel VLookUp* function. Matching individuals with unique alphanumerical codes was validated by examining additional identification information such as: *sex, occupation, place of residence, religion*.

The unique identification of individuals allowed us to interconnect the data from different respondents to build a multi-layered network. As illustrated in Figure 2, firstly, we generated *the personal network* of each of the 303 participants in the study. Each respondent (ego) is marked by a *trapezoid-shaped, black-bordered* node. Node color represents the place of living: "red" for Castellón (Spain), or "blue" for Dâmbovița (Romania) – *see* Figures 2A and 2B. Each ego is embedded in a personal network comprising a maximum number of 40 alters. Alters are designated by nodes of variant shapes and colors based on their corresponding type (class). Particularly, relatives are indicated by *squares*, friends by *triangles*, and acquaintances by *circles.* Colors again mark places of residence: "red" for Castellón (Spain), "blue" for Dâmbovița (Romania), "yellow" for other places (or countries) than Castellón or Dâmbovița. Ego's alters (relatives, friends and acquaintances) who had lived in Castellón and returned to live in Romania are marked by blue nodes with 'red' borders (*e.g.,* returned relatives are marked by blue squared nodes with red borders). Additionally, from the set of elicited alters a sub-set of alters were randomly sampled. The ties among these alters are marked by green edges.

In a second step, all of the 303 personal networks were interconnected through the link-tracing referrals. This is illustrated in Figure 2 by the black thick arrows. By *zooming in* into the visualization exhibited in Figure 2A, Figure 2B allows for a quick inspection of the interconnecting procedure. Each arrow has the referee as origin, while the arrow-head indicates the referral. Where applicable, the respondents' personal networks were also interconnected through their shared alters (common social contacts). Orange edges are indicative of alters shared by multiple respondents.

The sample from the TSF or the *network of networks* (the network generated by interconnecting personal networks, *see* Figure 2) is indicative of a multi-layered data structure. The first layer consists of 303 *personal networks* that can be independently analyzed, both in terms of composition and structural features. The second layer includes *the link-tracing network*, *i.e.,* the network that illustrates the implementation of the link-tracing sampling method. This network consists of 1,068 nodes and 1,187 ties. The third layer results from interconnecting personal networks (*the network of networks* that includes 4,855 nodes, 5,477 directed ties and 2,540 edges).

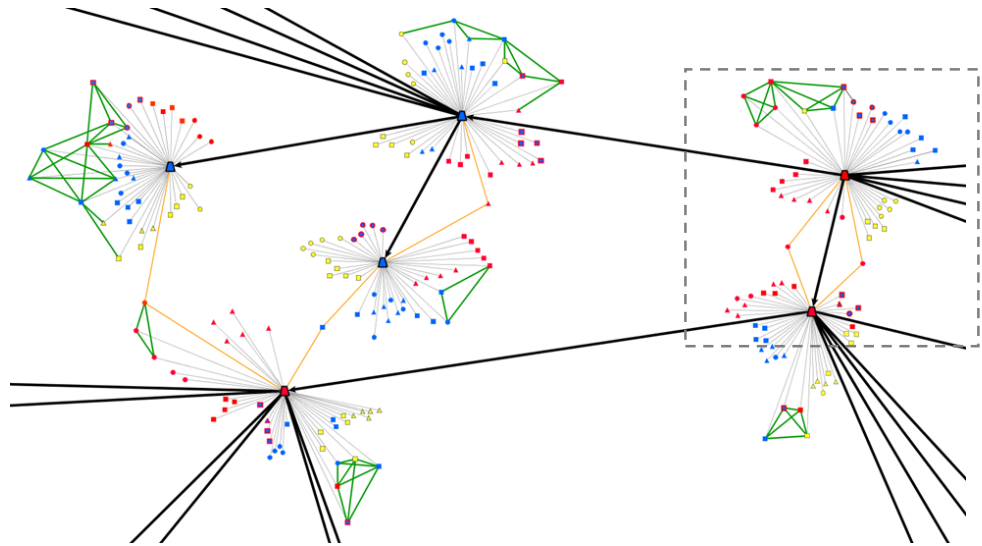

A. Inter-connected personal networks, through link-tracing ties and shared alters

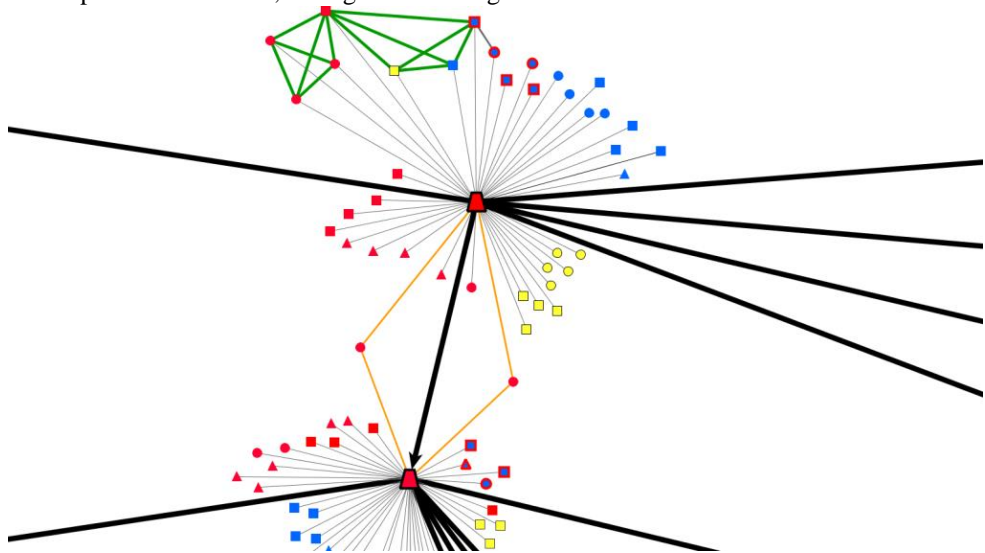

B. Zoom in on the personal network of respondents (the dotted area of A in the panel)

**Figure 2:** Multi-layered network (the network of networks) built by the link-tracing sampling method. *Note*: Node colors designate places (countries) wherein individuals currently live, *i.e.,* 'blue' for Dâmbovița (Romania), 'red" for Castellón (Spain), 'yellow' for other places than Dâmbovița and Castellón (these other places could be other regions in Romania and Spain, or even other countries – for simplicity and illustrative purposes, in this example, we decided, by 'yellow', to mark other countries. Node shapes designate classes of individuals (alters) elicited by the fixed-number alter name-generators: 'squares' designate ego's family members, 'triangles' designate ego's friends, while 'circles' designate ego's acquaintances. The ego (the respondent) is marked by a trapezoid shaped - black bordered node. Ego's alters (family members, friends and acquaintances) who lived in Castellón (Spain) and returned to live in Romania are marked by 'red' bordered shapes (*e.g.,* returned family members are marked by blue squared nodes with red border). A sub-set of nine alters was randomly sampled from the elicited set of alters, and the alter-alter existing ties represented. This is illustrated by the green edges. If two nodes are connected by an edge, that means the two nodes know each other and can contact each other independently from the ego. As all the nodes represent ego's alters, the respondent (the trapezoid shaped node) is connected to everybody through a tie (see the thin gray edges). Orange edges indicate the cases wherein two egos share alters. The black arrows indicate the direction of the link-tracing sampling: an arrow's origin marks the referee while the head of the arrow marks the referral. The plots were built using visone (Brandes and Wagner 2004).

## 3.5 The specificity of the ORBITS methodology

In this sub-section, we present the specificity of our current methodology (*i.e.,* the research project ORBITS: *The Role of Social Transnational Fields in the Emergence, Maintenance and Decay of Ethnic and Demographic Enclaves*) as well as the commonalities with the research of Mouw et al. 2014, Mouw and Verdery 2012 and Verdery et al. 2018 that was built on a similar binational link-tracing approach (*i.e., the Network Survey of Immigration and Transnationalism,* the *NSIT study*).

Firstly, the ORBITS study employed a simultaneously multi-sited data collection process, whereas the NSIT study collected the information in a temporal two-step manner (research in the destination places of Mexican migrants preceded research in their origin place). Secondly, the NSIT outsourced the data collection to community members while the ORBITS data was collected by the research team. Thirdly, the NSIT name generators (*i.e.* questions about participants' social contacts) were designed to collect information on maximum 27 personal contacts (alters) in the destination and 12 alters in the origin place. The ORBITS name generators allowed for collecting data on a maximum 40 personal contacts in each of the two sites (destination and origin places). Fourthly, there are two unique features of the ORBITS study; namely, (a) the collection of information on the ties between the personal contacts (alters) of the participants, *i.e.,* alter- alter edges (that provides insights on the structure of the personal networks of the participants), and (b) allowing respondents to also nominate people in other places which unveils participants' connections, beyond the migration corridor, to other countries worldwide. Table 1 illustrates in more detail the methodological differences between the two studies. In terms of commonalities, both studies: a) share the same research approach (binational link-tracing sampling from TSFs), b) collect data from the destination and the origin places of migrants (community-oriented procedure), c) collect cross-sectional data, and d) use demographic variables to uniquely identify TSF members, *i.e.* participants to the research as well as their nominees, either referrals or alters (as a way of ensuring both confidentiality and anonymity).

In the remainder of this paper, we describe the *participants' personal networks*, *the link-tracing network* and the *network of networks* (the network resulted from the interconnection of the personal networks of the participants to the study).

| Dimension | NSIT study | ORBITS study |
|---|---|---|
| Size of samples | Origin place: Guanajuato (Mexic), n = 407 (410) *<br>Destination places: North Carolina (US), n = 150 (146) and Houston (US), n = 52 (51) | Origin place: Dambovita (Romania), n = 154<br>Destination place: Castellon (Spain), n = 149 |
| The design of the data collection process | *Two-steps*: firstly, in the origin places, afterwards, in the destination.<br><br>(a) link-tracing sampling design, in the destination places: 12 seeds (North Carolina) & five seeds (Houston). The collection of data through link-tracing sampling was limited to the population of interest living in the destination places.<br><br>(b) "*pyramid selection approach*" (Mouw et al., 2014) or 4-level collection strategy in the origin place. 20 seeds (Guanajuato) were randomly selected from the pool of alters elicited in the first step (in the destination). On the 2-level, two friends and two relatives of each of the seeds were interviewed. On the 3-level, one friend and one relative of each of the participants on the 2-level were interviewed. On the 4-level, for each of the participants on the 3-level, one either friend or relative, randomly selected, was interviewed. | Link-tracing sampling design: nine seeds (Castellon).<br><br>From the nine seeds, the data collection process went on, being employed a simultaneously multi-sited data collection process.<br><br>Specifically, on the first wave, each seed was asked to nominate three referrals living in the origin and three, in the destination.<br><br>On the second wave, both the three referrals in the origin and the three referrals in the destination were asked to nominated three people in the origin and three in the destination.<br><br>Subsequently, on additional waves, referrals nominated by the referees interviewed in previous waves were also contacted and interviewed. The process halted when the sample size target was reached (at least 300 interviews). |
| Data collection | Community members collected the data, aside pretests. | Members of the ORBITS study team collected the data and conducted the pretests. ORBITS researchers simultaneously coordinated in a "*ping-pong*" fashion across the two sites. |
| Name generator | *Destination places*: ≤10 friends & ≤ 5 (6) relatives (living in the destination), ≤ 6 relatives / friends (living in the origin) and ≤ 5 returned migrants.<br>*Origin place*: ≤ 6 friends / relatives (living in the origin) and ≤ 6 friends / relatives living in the destination places.<br>Maximum unique number of elicited alters: 26(27) in the destination & 12 in the origin. | *Destination place*: ≤10 friends & ≤ 5 relatives (living in the destination), ≤ 5 returned to Romania friends & relatives, ≤ 5 relatives & ≤5 friends living in the origin, and ≤5 friends & ≤5 relatives living in other places (than the origin and the destination).<br>*Origin place*: ≤10 friends & ≤ 5 relatives (living in the origin), ≤ 5 returned to Romania friends & relatives, ≤ 5 relatives & ≤5 friends living in the destination, and ≤5 friends & ≤5 relatives living in other places (than the origin and the destination).<br>Maximum unique number of elicited alters: 40 (in both sites). |
| Alter-alter edges | | To avoid respondent burden, a sample of nine alters is randomly sampled from the pool of elicited alters by each participant. Subsequently, the alter-alter ties are measured, as existent or non-existent, according to the participants' perceptions. |

**Table 1**: Methodological differences between NSIT and ORBITS studies. *Note*: *The values in the parentheses, reported by Mouw et al (2014), are slightly different compared to the initially values reported by Mouw and Verdery (2012).

# 4 Results

## 4.1 Demographics of participants and refusals

Despite the demographic heterogeneity of the seeds (illustrated in Table 2), the initial sample was composed of individuals whose living experience in *Castellón* was consistent (on average they had been living in *Castellón* for 17 years). This was indicative of their high-level embeddedness in the local (*Castellón*) community as well as in the Romanian collectivity living in *Castellón*. This degree of embeddedness was deemed essential for starting and ensuring the success of the link-tracing sampling procedure.

| | |
|---|---|
| *Initial seeds (sample)* | 9 |
| *Sex* | |
| Males | 5 |
| Females | 4 |
| *Marital status* | |
| Married | 5 |
| Divorced & single | 1 |
| Widow(er) & single | 1 |
| Single | 2 |
| *Age* | |
| Mean (SD) | 44.4 (11.7) |
| Min (Max) | 27 (59) |
| *Level of formal education* | |
| Ten years completed with diploma | 1 |
| High school (with diploma) | 2 |
| Post high school education | 2 |
| Higher education (BA degree) | 4 |
| *Religion* | |
| Orthodox | 6 |
| Pentecostal | 1 |
| Adventist | 1 |
| No religion | 1 |
| *Work status* | |
| Employed | 8 |
| Unemployed | 1 |

**Table 2:** The demographic profile of the initial subjects in the sample (the seeds).

The link-tracing sampling method started from a number of nine seeds and continued, one wave after another, with a pile of 294 additional interviewees. The data collection process was stopped after the target sample size of at least 300 interviews had been reached (in our case, the process halted at 303 interviews). In the study, 1,059 referrals had been nominated in both two sites (not counting here the initial sample of nine seeds). As a general tendency, respondents tended to nominate more women (59%) than men (41%) ($\chi^2(1) = 28,61$, $p<.001$). Additionally, in Spain, 67% of all referrals were females while in Romania, 51%. Out of the total of 1,068 nodes comprising the link-tracing network, 765 people refused to participate to the study (72%) while 12 interviewees did not provide any referrals. It follows that the link-tracing sampling procedure had a success rate of nearly 28%.

Table 3 reports the distributions of refusals and participants, split by referee's gender and country of residence (place of living). The number of contacted people is about the same in *Romania* (539) as in *Spain* (529). In addition, in Dâmbovița, the participation to the study is roughly gender-balanced (53% males interviewed), compared to Castellón, wherein more than two thirds of participants were females (72%).

|  |  | *Participated in the research?* |  |  |
|---|---|---|---|---|
|  |  | Yes | No | Total |
| *Dâmbovița (Romania)* |  |  |  |  |
|  | Males | 83 (31%) <br> 53% | 180 (69%) <br> 47% | 262 (100%) <br> 49% |
|  | Females | 73 (26%) <br> 47% | 203 (74%) <br> 53% | 275 (100%) <br> 51% |
|  | Total | 156 (29%) <br> 100% | 383 (71%) <br> 100% | 539 (100%) <br> 100% |
| *Castellón (Spain)* |  |  |  |  |
|  | Males | 40 (27%) <br> 28% | 133 (76%) <br> 35% | 174 (100%) <br> 33% |
|  | Females | 107 (30%) <br> 72% | 249 (70%) <br> 65% | 358 (100%) <br> 67% |
|  | Total | 147 (28%) <br> 100% | 382 (72%) <br> 100% | 529 (100%) <br> 100% |
| *Total (both sites)* |  |  |  |  |
|  | Males | 123 (28%) <br> 40% | 313 (72%) <br> 41% | 436 (100%) <br> 41% |
|  | Females | 180 (28%) <br> 60% | 452 (72%) <br> 59% | 632 (100%) <br> 59% |
|  | Grand total | 303 (28%) <br> 100% | 765 (72%) <br> 100% | 1,068 (100%) <br> 100% |

**Table 3:** Distribution of participation by referee's gender and place of living. *Note:* The valid percentages are computed both column and row-wise; these may add up more than 100%.

The distributions for the main demographic characteristics of the participants to the ORBITS study can be inspected in Table 4. In Dâmbovița (Romania), 88% *never migrated to Spain*, while the average age within the site is *37-year-old.* Moreover, most of the participants in Romania were at least *high-school graduates* (36%) and *orthodox* (97%). In terms of *civil status, work status* and *parenthood*, the data indicate rather bi-modal distributions, *i.e., married* or *single*, *employed* or *inactive (students)*. In the place of destination (Castellón), the great majority of the interviewees were *employed* (59%; a much larger percentage than in Romania), *orthodox* (82%, fewer than in Romania), *parents* (69%, more than in Romania) and *high-school graduates or higher* (51%). They were *on average 44 years old.*

The link-tracing sample deviates, to various degrees, from the structure of the populations of Dâmbovița (Romania) and Castellón (Spain). Within the Romanian sub-sample, female participants represent 53%, whereas in population (18+ years old) their share is of 51%. The average age in the sample is lower (37) compared to the population (48). In the Spanish data-set, female proportion is 72%, which is way higher compared to their reported share in population, 53%. However, the average age is roughly the same, 43.5 to 42.0.

|  | Dâmbovița (Romania) *place of origin* | Castellón (Spain) *place of destination* | *Grand total* |
|---|---|---|---|
| *Respondents* | 156 (100%) | 147 (100%) | 303 (100%) |
| *Type of participants* | | | |
| Non-migrants | 138 (88%) | 0 (0%) | 138 (46%) |
| Return migrants | 18 (12%) | 0 (0%) | 18 (6%) |
| Migrants in Spain | 0 (0%) | 147 (100%) | 147 (48%) |
| *Sex* | | | |
| Female | 73 (47%) | 107 (72%) | 180 (60%) |
| Male | 83 (53%) | 40 (28%) | 123 (40%) |
| *Civil status* | | | |
| Married | 58 (37%) | 66 (45%) | 124 (41%) |
| Single (never married) | 57 (37%) | 36 (25%) | 93 (31%) |
| Unmarried & in a stable relationship | 26 (17%) | 8 (5%) | 34 (11%) |
| Divorced & single | 8 (5%) | 25 (17%) | 33 (11%) |
| Widow(er) & single | 7 (5%) | 9 (6%) | 16 (5%) |
| Separated & single | 0 (0%) | 1 (1%) | 1 (0%) |
| Other | 0 (0%) | 2 (1%) | 2 (1%) |
| *Age total* | | | |
| Mean (SD) | 37.2 (17.0) (*n = 156*) | 43,5 (13.5) (*n = 146*) | 40,2 (15.7) (*n = 302*) |
| Min (Max) | 19 (76) | 20 (73) | 19 (76) |
| *Age Females* | | | |
| Mean (SD) | 41.7 (18.6) (*n = 73*) | 44,4 (13.7) (*n = 106*) | 43,3 (15.9) (*n = 179*) |
| Min (Max) | 19 (76) | 21 (73) | 19 (76) |
| *Age Males* | | | |
| Mean (SD) | 33.2 (14.4) (*n = 83*) | 41,1 (12.6) (*n = 40*) | 35,8 (14.3) (*n = 123*) |
| Min (Max) | 19 (72) | 20 (63) | 19 (72) |
| *Level of formal education* | | | |
| No formal education | 0 (0%) | 1 (1%) | 1 (0%) |
| Less than four years | 0 (0%) | 0 (0%) | 0 (0%) |
| Four years completed | 0 (0%) | 0 (0%) | 0 (0%) |
| Between five and eight years | 3 (2%) | 1 (1%) | 4 (1%) |
| Eight years completed with certificate | 25 (16%) | 7 (5%) | 32 (11%) |
| Ten years completed with diploma | 12 (8%) | 32 (22%) | 44 (15%) |
| High school (without diploma) | 24 (15%) | 14 (10%) | 38 (13%) |
| High school (with diploma) | 56 (36%) | 50 (35%) | 106 (35%) |
| Post high school education | 10 (6%) | 19 (13%) | 29 (10%) |
| Higher education (BA degree) | 19 (12%) | 20 (14%) | 39 (13%) |
| Higher education (MA degree, PhD etc.) | 7 (5%) | 1 (1%) | 8 (3%) |
| *Work status* | | | |
| Employed | 59 (38%) | 86 (59%) | 145 (48%) |
| Self-employed | 10 (7%) | 10 (7%) | 20 (7%) |
| Unemployed | 1 (1%) | 18 (12%) | 19 (6%) |
| Retired | 17 (11%) | 9 (6%) | 26 (9%) |
| Student | 54 (35%) | 7 (5%) | 61 (20%) |

|  | Inactive | 14 (9%) | 7 (5%) | 21 (7%) |
|---|---|---|---|---|
|  | Other | 0 (0%) | 10 (7%) | 10 (3%) |
| *Religion* |  |  |  |  |
|  | Orthodox | 151 (97%) | 119 (82%) | 270 (90%) |
|  | Reformed | 0 (0%) | 0 (0%) | 0 (0%) |
|  | Pentecostal | 1 (1%) | 4 (3%) | 5 (2%) |
|  | Baptist | 0 (0%) | 7 (5%) | 7 (2%) |
|  | Adventist | 0 (0%) | 5 (3%) | 5 (2%) |
|  | Catholic | 0 (0%) | 0 (0%) | 0 (0%) |
|  | Other | 2 (1%) | 0 (0%) | 2 (1%) |
|  | No religion | 2 (1%) | 10 (7%) | 12 (4%) |
| *Do you have children?* |  |  |  |  |
|  | Yes | 73 (47%) | 101 (69%) | 174 (57%) |
|  | No | 83 (53%) | 46 (31%) | 129 (43%) |

**Table 4:** Major demographic characteristics of the link-tracing network participants (egos). *Note:* The valid percentages are computed column-wise; columns may add up more than 100%.

### 4.2 The link-tracing network

The link-tracing network (the network embedding 303 participants and another 765 referrals who did not participate to the study) consists of 1,068 nodes and 1,187 ties (see Figure 3). The hairball network allows for the inspection of the binational characteristic of the employed sampling design. Particularly, there are nine seeds (*red-colored down-triangles*) and 138 nodes (*red-colored circles)* that represent respondents living in Castellón, Spain. Additionally, *the blue-colored circles* represent interviewees in Dâmbovița, Romania. The pattern of providing referrals (the out-degree) is shown by proportionally increasing the size of each node. Supplementary, the direction of the link-tracing sampling is marked by directed ties.

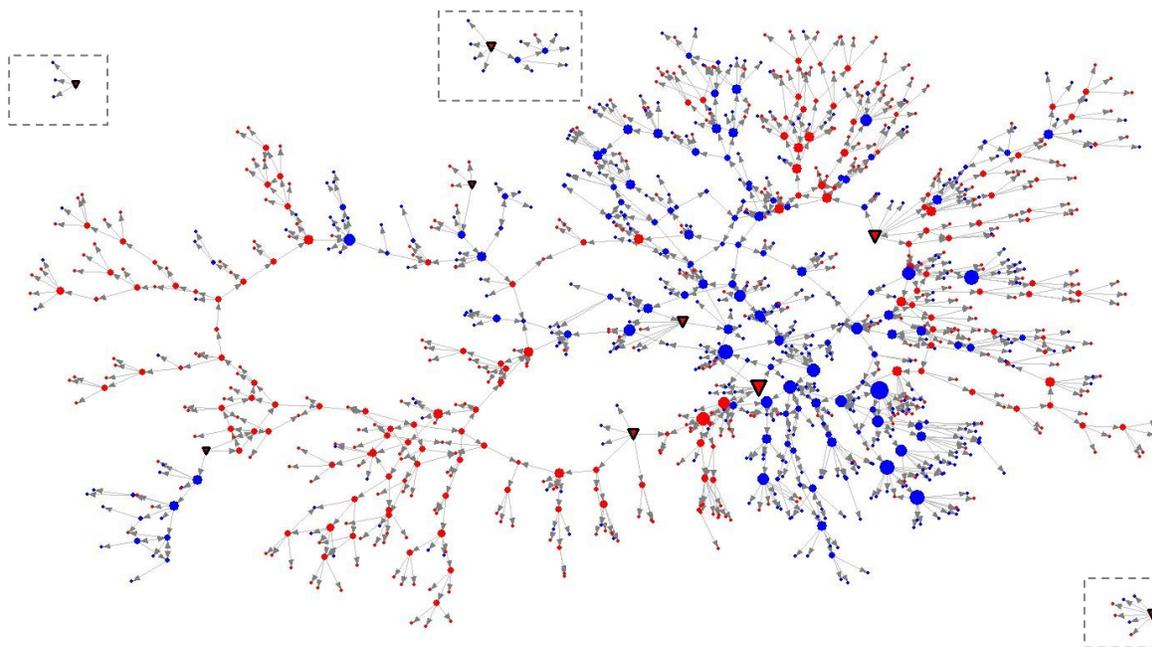

**Figure 3:** Hairball visualization of the link-tracing network. *Note:* Node colors indicate three classes of nodes: *red down-triangles* – seeds (individuals living in Castellón, Spain), *red* – people interviewed in Castellón, Spain, *blue* – people interviewed in Dâmbovița, Romania. The network is directed indicating the chain-referral structure of the relational data. Directed dyads are indicative of *referees* (the origin of the arrow) and referrals (indicated by the arrow head). Size nodes are proportional to their out-degree (the number of referrals provided). By dotted perimeters are indicated components (disconnected parts of the network). The plot was built using the UCINET 6.0 (Borgatti, Everett and Freeman 2002).

The network shows that six of the nine seeds, even though are not directly connected among each other, are embedded in the same *component* (a connected graph wherein all pairs of nodes are reachable through a succession of ties). Within this main component (accounting for 96% of all nodes), paths from some nodes to others are very large (with lengths up to 18). Given small world theories, it is likely that we have missed nodes that connect several of these cases. A reason for concern, as illustrated in Figure 3, is that in some chains, referrals to people in the other fieldwork site were either not given or not willing to participate, resulting in a clustering of country (color) in the network. The total 1,187 referee-referral arcs of the link-tracing network have the following distribution, within and across sites: *Romania – Romania* (*n = 421*), *Spain – Spain* (*n = 506*), *Romania – Spain* (*n = 97*) and *Spain – Romania* (*163*).

### 4.3 Referral pattern by sex and residence in the link-tracing network: Homophily

For all the referees and referrals in the link-tracing network (even for unsuccessful referrals), we had information about *sex* and *place of residence*. This information allows us to assess whether patterns of homophily can be identified. Specifically, on one hand, we could examine *intra-place* nominations (the tendency of participants to nominate referrals within their own place of residence) and *inter-place* nominations (the tendency toward cross-nomination by place of residence, *e.g.,* participants in Spain *(Castellón)* tend to rather nominate referrals in *Romania (Dâmbovița))*. On the other hand, we could inspect whether there are intra- or inter-sex nominations (*e.g.,* males tend to nominate males or males tend to nominate females). One way to explore response patterns is to work with the observed scores, without performing any further adjustments and manipulations. This naïve approach (Merli et al. 2016: 192) is suggested here only for illustrative purposes.

| from | | Females Romania | Males Romania | Females Spain | Males Spain |
|---|---|---|---|---|---|
| Spain | Males | 20 | 28 | 58 | 75 |
| Spain | Females | 77 | 38 | 294 | 79 |
| Romania | Males | 63 | 166 | 32 | 23 |
| Romania | Females | 135 | 57 | 27 | 15 |

**to**

**Figure 4:** Pattern of referral nominations based on sex and country residence. *Note*: The visualization illustrates the pattern of referral nominations, based on sex and place of residence. For instance, the female respondents living in Romania (Dâmbovița) (see the "from" axis) indicated as referrals (see the "to" axis): 135 and 57 males living in Romania (Dâmbovița), 27 females and 15 males living in Spain (Castellón).

Figure 4 shows the distribution of observed scores over *sex* and *place of residence*. *Homophily* and *heterophily sex effects* are indicated either in association with the place of residence, or independently. The color of each cell variates in intensity as a function of the frequency. Figure 4 illustrates a census of all the nominations. The four-by-four matrix is a mix of sex and place of residence (country). It indicates, for instance, the nomination tendency of Romanian females; the rows of the matrix are the referees, while the columns are the referrals. The two-by-two matrix allows for independently inspecting male and female-homophily effect. It should be noted that: a) there is an overall tendency toward male- and female-homophily; b) the sex nomination pattern is affected by the place of residence, *e.g.,* male respondents tend to rather nominate more males living in their proximity (country) rather in the other site (that also holds for female nomination pattern).

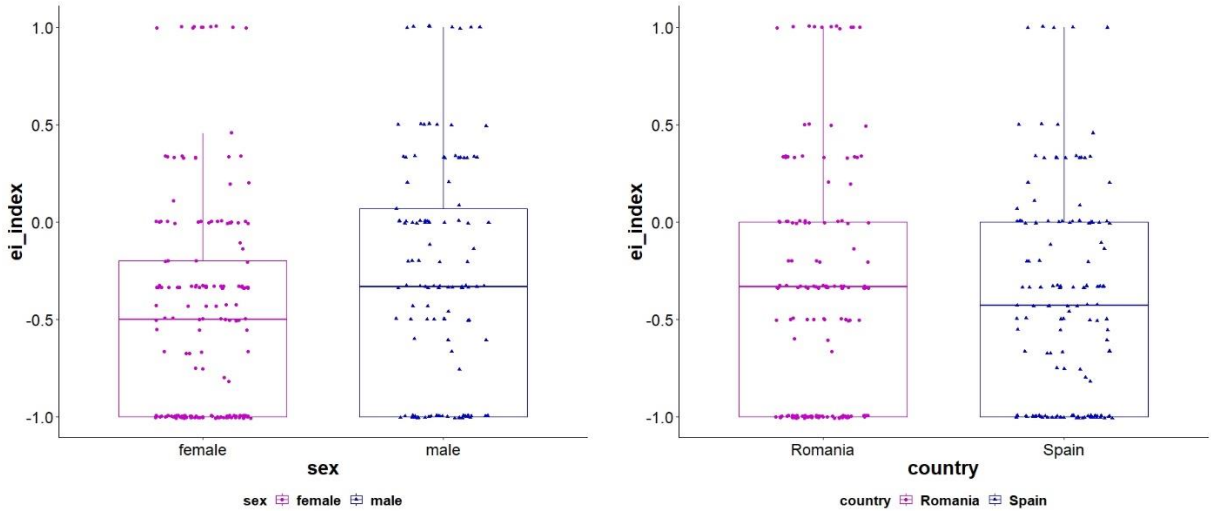

**Figure 5**: Boxplots illustrating individual distributions of EI-index scores split on sex and country of residence. *Note*: For every study participant who provided referrals, an E-I index score was computed, either taking into account sex or place of residence (country). E-I index is computed by a simple formula: for a specific participant the difference between the number of ties between groups and within group is divided up to the sum of ties (*e.g.,* for a female participant, the difference between her ties sent to males and her ties sent to other females is divided up to the total number of ties).

Figure 5 illustrates the univariate distribution of E-I index (Krackhardt and Stern 1988) scores computed independently for sex and place of residence, in UCINET 6.0 (Borgatti, Everett, and Freeman 2002). The index ranges from -1.0 to +1.0 where positive scores indicate heterophily (*e.g.,* the tendency of males referring to females, of people in Romania to people in Spain). Negative E-I index scores indicate homophily. Figure 5 shows that throughout the link-tracing network there are both sex and place of residence tendencies toward homophily.

|  | E-I index scores | | |
| --- | --- | --- | --- |
|  | Expected | SD | Observed |
| Sex | -0.033 | 0.031 | -0.391* |
| Country of residence | 0.001 | 0.029 | -0.561* |

**Table 5**: Sex / place of residence E-I index scores computed in the link-tracing network. *Note*: The scores were computed using the *E-I* index routine available in UCINET 6.0 (Borgatti, Everett and Freman 2002). Both observed E-I index scores on *sex* and *country of residence* were statistically significant ($p < .05$). The expected *E-I* index scores were computed after conducting 5,000 permutations. The computations took into account the link-tracing size (n= 1,068) and a total edge number of 1,187.

Working with the observed values (frequencies) for detecting homophily has two potential validity threats (Merli et al. 2016): nominations are conditioned by the relative number of people in each category (*e.g.,* number of females versus number of males, respondents living in Romania versus respondents living in Spain) and by the volume of referrals (*i.e.,* the number of referrals variates across referees). To manage these potential validity threats, we performed permutation tests to assess whether the observed link-tracing network E-I index score on a specific attribute (sex and residence) is significantly higher or lower than the score expected by chance. According to the results reported in Table 5, the *sex* and *place of residence* uniform homophily effects still hold.

We also used Exponential Random Graph Modeling (ERGM), *i.e.*, statistical models that allow for explaining tie patterning. Basically, tie formation processes within social networks can be accounted for by looking at: a) various local network configurations (*e.g.*, transitivity structures - *friends of friends are friends,* reciprocity - *you scratch my back and I scratch yours*, preferential attachment, etc.); b) actor attributes *or* node covariates (*e.g.,* homophily); c) dyadic covariates (geographical distance between persons) (Lusher, Koskinen, and Robins 2013).

To examine the probability of a tie to exists in the link-tracing network, we built several ERGM models (Handcock et al. 2003). We assessed several determinants of the nomination tie pattern within the link-tracing network. Specifically, we looked at: the propensity of people to make nominations (*sociality*), the propensity to make nominations based on attributes such as country residence and sex (*assortative mixing*) and based on *triad-based clustering* (the tendency of two link-tracing network members, either they are connected or not, to share a third member). Sociality, assortative mixing and triad-based clustering are expected to produce within the network, *homophily*, *transitivity* and a specific *degree distribution* (the distribution of nomination ties) (Goodreau, Kitts, and Morris 2009) .

Precisely, in terms of model specification, our inferences on sociality were based on counting observed ties (number of nominations and number of times being nominated). Additionally, we examined *uniform homophily* (overall propensity to make nominations based on country residence and sex) and *differential homophily* (*e.g.*, the propensity of males / females to nominate similar others) (Morris, Handcock, and Hunter 2008). For triad-based clustering, we implemented *Geometrically weighted dyad wise shared partner distribution* (GWDSP) (Hunter 2007). The models were fit on the link-tracing network ($N = 1,068$ individuals, see Figure 3) as well as on the link-tracing network of participants only ($N = 303$, the respondents). The link-tracing network of participants resulted from filtering out the 765 referrals who did not participate to the study. Given that *the link-tracing network of participants* represents less than a third from the entire link-tracing network (of nominated people), we were interested in exploring potential differences in terms of micro-level determinants of the nomination ties.

Table 6 displays the results of the ERGM models fitted to both networks. We report the conditional log-odds of two actors being connected by a nomination tie (either $i \rightarrow j$ or $i \leftrightarrow j$), *i.e.*, the estimates of the micro-level determinants. In all four models, the sociality coefficient (*edges*) is negative and statistically significant, indicating low density. Uniform homophily is reported in the link-tracing network of participants (Model 3). Specifically, the coefficients of the predictors ($\theta$ *coefficients)* indicate more *sex* and *country homophily* than expected by chance, *i.e.* there is a tendency for within *sex* and *country* categories edges. For instance, there is .58 more chances (*i.e., exp*(0.341) / (1+*exp*(0.341))) to see *a male nominating a male* or *a female nominating a female* than we would have in a random network.

In terms of differential homophily, *male - male* and *female – female* nomination effects are positive and statistically significant in the link-tracing of participants (Model 4). Also, *within country nomination* patterns are reported to be positive and significant (Model 4). These indicate that there is a general tendency of ORBITS study participants to rather nominate people of *the same sex* and living within *the same country*. The homophily effects are not statistically significant in the link-tracing network (Model 1 and 2). It is possible that the results (the effects) were affected (or filtered out) by the high number of network non-participants (as mentioned earlier, the link-tracing network is composed of 72% of non-respondents). Lastly, the tendency of two network members, irrespective of whether they are connected or not, to share a third member is statistically significant only in the link-tracing network (Models 1 and 2). This triad-based clustering effect is not significant for the link-tracing network of study participants (Models 3 and 4).

|  | Link-tracing network | | | | Link-tracing of participants | | | |
|---|---|---|---|---|---|---|---|---|
|  | Model 1 | | Model 2 | | Model 3 | | Model 4 | |
|  | Estimate | S.E. | Estimate | S.E. | Estimate | S.E. | Estimate | S.E. |
| ***sociality*** | | | | | | | | |
| edges | -7.149 *** | 0.079 | -7.139 *** | 0.080 | -5.820 *** | 0.151 | -5.812 *** | 0.154 |
| ***selective mixing*** | | | | | | | | |
| *uniform homophily* | | | | | | | | |
| sex | 0.035 | 0.057 | | | 0.341 ** | 0.107 | | |
| country | 0.034 | 0.057 | | | 0.456 *** | 0.108 | | |
| *differential homophily* | | | | | | | | |
| male - male ties | | | 0.078 | 0.080 | | | 0.417 ** | 0.141 |
| female - female ties | | | 0.004 | 0.061 | | | 0.288 * | 0.121 |
| Spain - Spain ties | | | 0.119 | 0.069 | | | 0.567 *** | 0.126 |
| Romania - Romania ties | | | -0.058 | 0.072 | | | 0.347 ** | 0.131 |
| ***triad-based clustering*** | | | | | | | | |
| GWDSP | 0.106 *** | 0.025 | 0.103 *** | 0.025 | -0.035 | 0.046 | -0.036 | 0.046 |
| AIC | 18666 | | 18665 | | 4914 | | 4915 | |
| BIC | 18714 | | 18737 | | 4951 | | 4972 | |

**Table 6**: Micro-level determinants of nomination tie patterns, in the link-tracing network. *Note:* \*\*\* *p<.000*, \*\* *p<.001*, \**p<.01*. The Monte Carlo MLE results were computed using the statnet suite (Handcock et al. 2003).

## 4.4 The link-tracing network chains

The link-tracing network can be decomposed into nine chains. These chains are *referee – referral waves* or succession of nodes (both participants and non-participants) stemming from each of the nine seeds. It should be stressed that a node may appear in more than one chain. This explains why six of the nine chains collapse into one main component (see Figure 3) and, in effect, why the total number of nodes embedded in each chain exceeds the number of nodes within the link-tracing network. Table 7 displays the demographic composition of these network chains as well as, for comparative purposes, of the link-tracing network. Each of the network chains is assigned its corresponding seed. Several things can be noticed in Table 7. Firstly, approximately 61% of the total link-tracing network nodes (*i.e.,* 656) and more than a third of the total participants (117 out of 303) are embedded in the chain stemming from seed 2. Secondly, chains variate in composition of the country wherein referrals currently reside. For instance, the chains stemming from the second and sixth seeds are dominated by people with residence in Spain, whereas chains stemming from the third and fifth seeds, by people in Romania. Thirdly, a gender variation across chains can be noticed, *e.g.,* the second seed chain has 61% females, whilst the third seed chain, 51% males. Fourthly, across chains and link-tracing network, the average distance of any pair of nodes variates between one and six.

The variations of chains in terms of residence or sex may indicate homophily governing the tie nomination patterns. Table 8 reports the results of ERGM models fitted to the largest three chains (*i.e.,* chains stemming from the second, third and sixth seed). The estimates show in all the three models that there are less nominations than we would expect by chance alone (negative, statistically significant $\theta$ *coefficients*). In the chain with the largest number of nodes (from seed 2), homophily was not detected; the estimates of differential homophily on *sex* and *country* were not statistically significant. However, a *male-male homophily effect* was detected in the third seed's chain ($\theta = .514$, $p < .01$) and a *Spain-Spain homophily effect* was detected in the sixth seed's chain ($\theta = -.328$, $p < .001$).

To visualize the variation in structure and composition of referral chains by seed, Figure 6 provides a panel of hive-plots (Hanson 2017) both for the link-tracing network and for all the nine chains. These visualizations display the pattern of nominations within and between the two sites (Romania and Spain). Hive-plots axes are unstandardized and proportional to the nodes assigned to the two sites (this allows for visually inspecting whether in a specific chain there are more people living in Spain or Romania). For instance, the *Seed 2 network* contains more people living in Spain, compared to the *Seed 3 network* which is dominated by people living in Romania. For some seed networks (1, 4, 8 and 9), axes and nomination ties are difficult to identify due to the low number of nodes (see Table 7 for their corresponding composition).

|  | Seed#1 | Seed#2 | Seed#3 | Seed#4 | Seed#5 | Seed#6 | Seed#7 | Seed#8 | Seed#9 | link tracing network |
|---|---|---|---|---|---|---|---|---|---|---|
| *Type of participants* | | | | | | | | | | |
| Migrants in Spain | 1 | 107 | 16 | 3 | 1 | 41 | 9 | 11 | 1 | 147 |
| Non-migrants | 0 | 56 | 64 | 0 | 11 | 10 | 1 | 0 | 0 | 138 |
| Return migrants | 0 | 14 | 3 | 0 | 0 | 0 | 2 | 0 | 0 | 18 |
| *Type of network members* | | | | | | | | | | |
| Participants | 1 (14%) | 177 (27%) | 83 (34%) | 3 (23%) | 12 (34%) | 51 (23%) | 12 (34%) | 11 (37%) | 1 (25%) | 303 (28%) |
| Non-participants | 6 (86%) | 479 (73%) | 163 (66%) | 10 (77%) | 23 (66%) | 166 (77%) | 23 (66%) | 19 (63%) | 3 (75%) | 765 (72%) |
| *Type of network members by country* | | | | | | | | | | |
| *Spain* | | | | | | | | | | |
| Participants | 1 (14%) | 107 (16%) | 16 (7%) | 3 (23%) | 1 (3%) | 41 (19%) | 9 (26%) | 11 (37%) | 1 (25%) | 147 (14%) |
| Non-participants | 3 (43%) | 264 (40%) | 45 (18%) | 9 (69%) | 9 (26%) | 117 (54%) | 14 (40%) | 17 (57%) | 3 (75%) | 382 (36%) |
| Total network members | 4 (57%) | 371 (56%) | 61 (25%) | 12 (92%) | 10 (29%) | 158 (73%) | 23 (66%) | 28 (93%) | 4 (100%) | 529 (50%) |
| *Romania* | | | | | | | | | | |
| Participants | 0 (0%) | 70 (11%) | 67 (27%) | 0 (0%) | 11 (32%) | 10 (4%) | 3 (9%) | 0 (0%) | 0 (0%) | 153 (14%) |
| Non-participants | 3 (43%) | 215 (33%) | 118 (48%) | 1 (8%) | 14 (40%) | 49 (23%) | 9 (26%) | 2 (7%) | 0 (0%) | 383 (36%) |
| Total network members | 3 (43%) | 285 (44%) | 185 (75%) | 1 (8%) | 25 (71%) | 59 (27%) | 12 (34%) | 2 (7%) | 0 (0%) | 539 (50%) |
| *Type of network members by sex* | | | | | | | | | | |
| Males | 0 (0%) | 257 (39%)* | 126*(51%) | 2 (15%) | 6 (17%) | 69 (32%)* | 12 (34%)* | 10 (33%) | 2 (50%)* | 436 (41%) |
| Females | 7 (100%)* | 399 (61%) | 120 (49%) | 11 (85%)* | 29 (83%)* | 148 (68%) | 23 (66%) | 20 (67%)* | 2 (50%) | 632 (59%) |
| *Type of network members by country and sex* | | | | | | | | | | |
| *Spain* | | | | | | | | | | |
| Male participants | 0 (0%) | 33 (5%) | 1 (0%) | 0 (0%) | 0 (0%) | 12 (6%) | 3 (9%) | 4 (13%) | 1 (25%) | 40 (4%) |
| Male non-participants | 0 (0%) | 94 (14%) | 20 (8%) | 2 (15%) | 1 (3%) | 34 (16%) | 4 (11%) | 5 (17%) | 1 (25%) | 133 (12%) |
| Female participants | 1 (14%) | 74 (11%) | 15 (6%) | 3 (23%) | 1 (3%) | 29 (13%) | 6 (17%) | 7 (23%) | 0 (0%) | 107 (10%) |
| Female non-participants | 3 (43%) | 170 (26%) | 25 (10%) | 7 (54%) | 8 (23%) | 83 (38%) | 10 (29%) | 12 (40%) | 2 (50%) | 249 (23%) |
| *Romania* | | | | | | | | | | |
| Male participants | 0 (0%) | 36 (5%) | 42 (17%) | 0 (0%) | 1 (3%) | 4 (2%) | 0 (0%) | 0 (0%) | 0 (0%) | 83 (8%) |
| Male non-participants | 0 (0%) | 94 (14%) | 63 (26%) | 0 (0%) | 4 (11%) | 19 (9%) | 5 (14%) | 1 (3%) | 0 (0%) | 180 (17%) |
| Female participants | 0 (0%) | 34 (5%) | 25 (10%) | 0 (0%) | 10 (29%) | 6 (3%) | 3 (9%) | 0 (0%) | 0 (0%) | 73 (7%) |
| Female non-participants | 3 (43%) | 121 (18%) | 55 (22%) | 1 (8%) | 10 (29%) | 30 (14%) | 4 (11%) | 1 (3%) | 0 (0%) | 203 (19%) |
| **Chain volume (# nodes)**\*\* | 7 (100%) | 656 (100%) | 246 (100%) | 13 (100%) | 35 (100%) | 217 (100%) | 35 (100%) | 30 (100%) | 4 (100%) | 1068 (100%) |
| *Longest distance from the seed* | 1 | 16 | 14 | 3 | 5 | 13 | 6 | 8 | 1 | - |
| *Average distance from the seed* | 1 | 7.9 | 8.6 | 1.8 | 2.7 | 6.3 | 3.5 | 3.0 | 1 | - |
| *Average distance in the network (SD)* | 1.0 (0.0) | 6.2 (3.5) | 5.9 (3.4) | 1.7 (0.8) | 2.1 (1.0) | 5.1 (2.9) | 2.6 (1.4) | 2.9 (1.9) | 1.0 (0.0) | 6.1 (3.5) |
| *Proportion in the link-tracing network ** * | .7% | 61.4% | 23.0% | 1.2% | 3.3% | 20.3% | 3.3% | 2.8% | .4% | - |

**Table 7:** The demographic composition of the link-tracing network chains. *Note:* Percentages are computed column-wise based on each chain's volume (number of nodes). In some cases, these may exceed 100%. * Within computations, seeds are included in the class of participants; ** Some of the participants appear in more than one seed-chain. For this reason, summation of chain volumes exceeds the total number of network members (1,068).

| | Seed 2 Model 1 | | | Seed 3 Model 2 | | | Seed 6 Model 3 | | |
|---|---|---|---|---|---|---|---|---|---|
| | Estimate | | S.E. | Estimate | | S.E. | Estimate | | S.E. |
| *sociality* | | | | | | | | | |
| edges | -4.963 | *** | 0.192 | -4.286 | *** | 0.307 | -5.549 | *** | 0.181 |
| *selective mixing differential homophily* | | | | | | | | | |
| *sex* | | | | | | | | | |
| male - male ties | -0.083 | | 0.205 | 0.514 | * | 0.253 | -0.193 | | 0.251 |
| female - female ties | 0.138 | | 0.144 | 0.130 | | 0.288 | 0.144 | | 0.125 |
| *country* | | | | | | | | | |
| Spain - Spain ties | 0.207 | | 0.146 | -0.122 | | 0.624 | -0.328 | ** | 0.126 |
| Romania - Romania ties | -0.094 | | 0.211 | -0.069 | | 0.259 | 0.194 | | 0.291 |
| *triad-based clustering* | | | | | | | | | |
| GWDSP | -0.014 | | 0.058 | 0.103 | | 0.099 | 0.172 | *** | 0.047 |
| AIC | 2752 | | | 1008 | | | 2999 | | |
| BIC | 2802 | | | 1049 | | | 3052 | | |

**Table 8:** Micro-level determinants of nomination tie patterns, in three chains. *Note:* *** $p<.000$, ** $p<.001$, * $p<.01$. The Monte Carlo MLE results were computed using the statnet suite (Handcock et al. 2003).

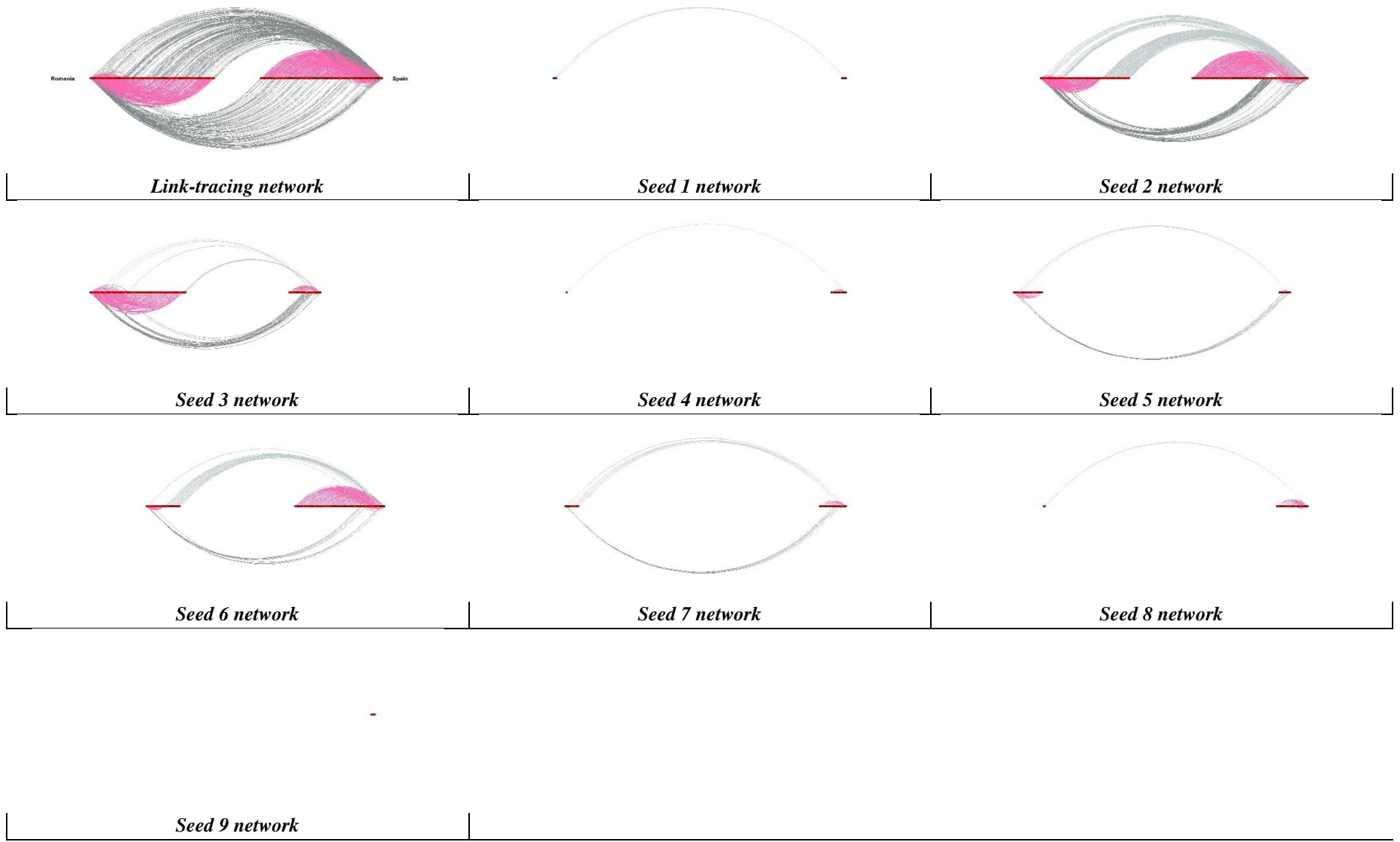

**Figure 6:** Hive-plots of the link-tracing network and of the seed chains, by residence. *Note*: The link-tracing network consists of 1,068 nodes that are placed on two axes, based on residence (either Castellón – Spain on the right, or Dâmbovița – Romania, on the left). The other nine hive-plots display the chains stemming

from each of the nine seeds. In the plots, each node is positioned on the axis based on its rank (out-degree or the number of people they nominated), under the principle of first served (for equal degrees, nodes are placed based on the assigned numbering in the dataset). Node colors are indicating each node's sex: dark-red indicates females, while red, males. The top ties indicate the nomination pattern of people interviewed in Spain: the magenta ties indicate referee – referral living in Castellón, whereas the gray lines indicate a referee from Castellón and a referral in Dâmbovița. The bottom ties indicate the nomination pattern of people interviewed in Romania: the magenta lines indicate referee – referral dyads living in Dâmbovița, while the gray ties indicate a referee from Dâmbovița and a referral in Castellón. The hiveplots were built using the hiveR package (Hanson 2017).

## 4.5 The network of networks

In the remainder of this section, we report results on the compositional and structural features of the *network of networks, i.e.,* the network built by interconnecting the link-tracing network and the personal networks of participants (specifically, their references to personal contacts, alters, other than the referrals). The network of networks consists of 4,855 nodes (participants, referrals, and alters), 5,477 arcs (nomination ties) and 2,540 symmetric ties (alter-alter ties). Figure 7 illustrates the network of networks using a hair-ball layout (*i.e., stress minimization node-layout* available with visone (Brandes and Wagner 2004)). Node colors mark the country of residence (*red* for Spain, *blue* for Romania, and *green* for other countries). Despite the large volume of nodes, four components can be identified, *i.e.,* a main (giant) component that accounts for 95% of all nodes, and three small components, accounting for the other 5%.

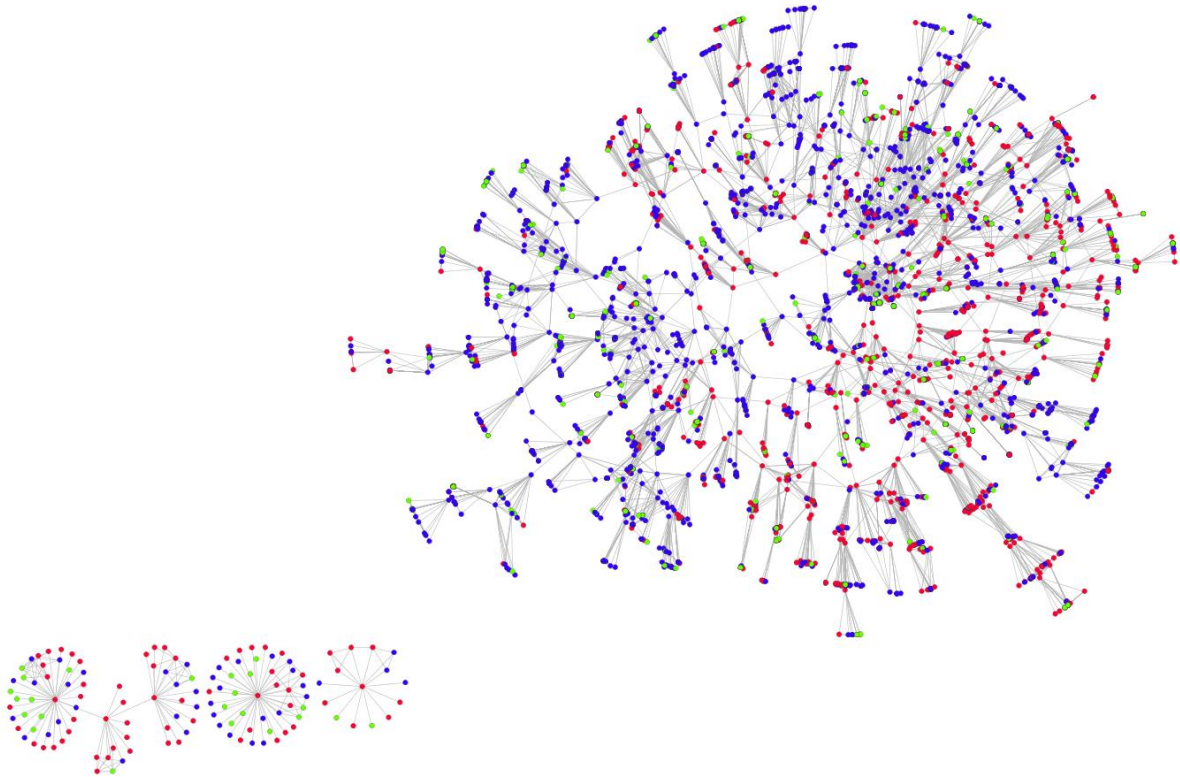

**Figure 7:** The network of networks. *Note:* The network has 4,855 nodes, 5,477 arcs and 2,540 undirected ties. Colors indicate network nodes' place of living, *i.e.,* red (Spain), blue (Romania), and green (other countries). The network data was visually encoded with visone (*stress minimization* node layout) (Brandes and Wagner 2004).

Figure 8 visually encodes the network of networks using a hive-plot format. Nodes are distributed on three axes based on their residence. Magenta ties indicate edges within the same country, *i.e.,* Spain – Spain (1,524 ties) or Romania – Romania (2,237). The axis illustrating the "*other* countries" class of nodes lacks within-ties by design. Nodes assigned to this axis were not interviewed, but only nominated by the participants (contacts they have in their personal network, either relatives, friends or acquaintances). The gray colored lines indicate inter-country ties. There are 1,133 ties connecting Spain and Romania. Also, there are 223 ties sent from Spain and 360 ties sent from Romania, to Romanians living in other countries. This visualization suggests that interviewees are not only a part of the binational corridor (Spain-Romania) but are also connected to other corridors. The axes are unstandardized to emphasize differences in the volumes of nodes. There are 1,656 nodes assigned to the "Spain" axis (1,049 females – 63%), 2,638 nodes to the "Romania" axis (1,337 females – 51%) and 561, to the "Other countries" axis (269 females

– 48%). On each axis, node placement was ranked by the out-degree (the number of nominations elicited by each participant), while for equal out-degree, placement was based on the indexation in the database.

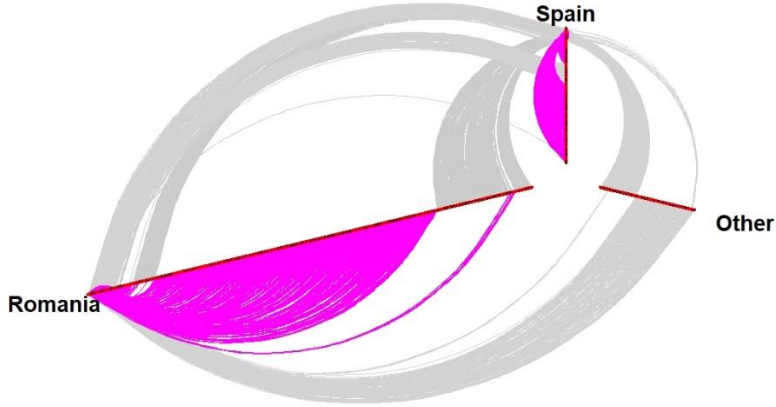

**Figure 8:** The network of networks (ties split on residence). *Note:* The hiveplot illustrates how the 4,855 nodes and the 5,477 nomination ties within the network of networks are partitioned on residence (countries wherein people live). Each node is positioned on the axis based on its rank (*out-degree* or the number of people they nominated), under the principle of *first served* (for equal degrees, nodes are placed based on the assigned numbering in the dataset). Node colors are indicating each node's sex: dark-red indicates females, while light red, males. Magenta ties indicate *within-country* social connections (*i.e.* 1,524 ties connect people living in Spain, and 2,237, people living in Romania), whereas gray ties, *between-country* connections (*i.e.* 1,133 ties connect Spain and Romania, 223 ties connect Romanians living in Spain to Romanians living in other countries, and 360 ties connect Romanians living in Romania to Romanians living in other countries). The hiveplot was built using the hiveR package (Hanson 2017).

|  | Migrants in Spain (n = 147) |  | Returned migrants (n = 18) |  | Non migrants (n = 138) |  | ORBITS study participants (n = 303) |  |
|---|---|---|---|---|---|---|---|---|
|  |  | Mean (SD) |  | Mean (SD) |  | Mean (SD) |  | Mean (SD) |
| Acquaintances and friends, living in Spain | 824 | 5.6 (2.3) | 29 | 1.6 (2.3) | 110 | 0.8 (1.3) | 963 | 3.2 (3.0) |
| Acquaintances and friends, living in Romania | 258 | 1.8 (2.2) | 136 | 7.6 (3.2) | 1138 | 8.2 (3.3) | 1,532 | 5.1 (4.3) |
| Acquaintances and friends, living in other countries | 113 | 0.8 (1.2) | 17 | 0.9 (1.8) | 152 | 1.1 (1.5) | 282 | 0.9 (1.4) |
| *Acquaintances and friends* | 1,195 | 8.1 (4.7) | 182 | 10.1 (3.8) | 1,400 | 10.1 (4.1) | 2,777 | 9.2 (4.5) |
| Family, living in Spain | 348 | 2.4 (1.6) | 38 | 2.1 (1.9) | 161 | 1.2 (1.7) | 547 | 1.8 (1.7) |
| Family, living in Romania | 387 | 2.6 (1.9) | 100 | 5.6 (2.5) | 719 | 5.2 (2.1) | 1,206 | 4.0 (2.4) |
| Family, living in other places | 103 | 0.7 (1.1) | 11 | 0.6 (1.3) | 172 | 1.2 (1.6) | 286 | 0.9 (1.4) |
| *Family* | 838 | 5.7 (3.4) | 149 | 8.3 (2.9) | 1,052 | 7.6 (3.1) | 2,039 | 6.7 (3.4) |
| ***Elicited alters*** | **2,033** | **13.8 (7.2)** | **331** | **18.4 (3.9)** | **2,452** | **17.8 (5.9)** | **4,816** | **15.9 (6.8)** |

**Table 9**: Number of alters nominated within the name generators, split by type of participants.

Table 9 reports the *alters* (people embedded in the personal networks of the participants) elicited by the respondents. The respondents elicited on average 16 alters (the maximum was 40), summing up to a total of 4,816. Inspecting the average scores across alter categories and respondent classes, several things are noteworthy. Firstly, migrants in Spain, as expected, nominated three times more acquaintances and friends living in Spain, compared to the other two respondent categories. Secondly, people with migration experience in Spain (either currently living there, or returnees) nominated more relatives living in Spain compared to non-migrants. Thirdly, migrants in Spain nominated almost half of the relatives and roughly four times less friends living in Romania compared to returnees and non-migrants. Fourthly, people with no migration experience in Spain (non-migrants) nominated more relatives (almost double) and slightly more friends living in other countries compared to the participants with Spanish migration experience. Fifthly, overall, migrants in Spain nominated fewer acquaintances and friends and relatives compared to the other two classes of respondents.

Table 10 presents a general overview on the structural features exhibited by the *network of networks* as well as by the *link-tracing network* and *chain-networks*. These features are displayed on three levels of measurements: basic elements, dyadic and network level. The network of networks has 4,855 nodes, 5,477 directed ties (nominations) and 162 mutual dyads, *i.e.,* referrals who nominated their referees (this number of mutual dyads is almost insignificant given that the total number of possible reciprocated dyads in the network is of more than 11 million; we excluded from computations the 2,540 undirected ties representing the alter-alter ties). The extremely low dyadic reciprocity can also be noticed in the other layers (the link-tracing and chain networks) of the network of networks.

Concerning the *network density* (observed by expected ties), the ratio decreases as the number of nodes increases (large networks are generally scarce in terms of ties). Moreover, as reported throughout the paper, all the ERGM models indicated that the number of observed ties is lower compared to the one expected by chance alone (*see* Tables 6 and 8). Looking at the results reported in the network level measurements section of the Table 10, we notice that, generally, the networks exhibit low levels of centralization. In a nutshell, network centralization would be high when nodes vary severely in degree distributions (some nodes would have significantly more ties compared to others - the "all roads lead to Rome" effect). The centralization for the *network of networks* is less than 1% irrespective of the centralization measurement (indegree, degree, degree or betweenness). As a general remark, for all the valid networks (networks that have a sufficient number of nodes so as the centralization measurements would have a substantial meaning) reported in Table 10, centralization scores do not exceed 20.7%. Specifically, across valid networks: a) *indegree centralization* (centralization that takes into account only the number of times a node gets nominated) variates between 2.0% and 11.1%; b) *outdegree centralization* (centralization that accounts only for the nominations made by a node) variates between 0.9% and 20.7%; c) *degree centralization* (centralization that take into account the number of times a node nominates and gets nominated by other nodes) variates between 0.5% and 10.3%; d) betweennness centralization (centralization that accounts for the number of times a node is placed, in the network, between other two nodes) variates between 0.2% and 4.1%.

The last network-level measurement reported in Table 10 concerns the number of components (in a component, all pairs of nodes are reachable). Nodes that compose the seed (chain) networks are by design part of the same component. The *network of networks* has four components (the main component includes 95% of the 4,855 nodes). The link-tracing networks also have four components: *the link-tracing network of participants* (wherein non-participants were filtered out) has a main component that includes 98% of the 303 nodes, whereas the *full link-tracing network* has a main component that includes 96% of the 1,068 nodes.

|  | Seed 1[a] | Seed 2 | Seed 3 | Seed 4[a] | Seed 5 | Seed 6 | Seed 7 | Seed 8 | Seed 9[a] | Linktracing | Linktracing[b] | Network of networks[d] |
|---|---|---|---|---|---|---|---|---|---|---|---|---|
| ***Basic elements*** | | | | | | | | | | | | |
| Nodes | 7 | 656 | 246 | 13 | 35 | 217 | 35 | 30 | 4 | 1,068 | 303 | 4,855 |
| Ties | 6 | 732 | 269 | 12 | 34 | 240 | 36 | 36 | 3 | 1,187 | 382 | 5,477 |
| ***Dyads*** | | | | | | | | | | | | |
| Mutual dyads | 0 | 17 | 2 | 0 | 0 | 5 | 2 | 4 | 0 | 24 | 24 | 162 |
| Asymmetric dyads | 6 | 698 | 265 | 12 | 34 | 230 | 32 | 28 | 3 | 1,139 | 334 | 5,153 |
| ***Network level measurements[d]*** | | | | | | | | | | | | |
| Network density | 0.143 | 0.002 | 0.004 | 0.077 | 0.029 | 0.005 | 0.030 | 0.041 | 0.250 | 0.001 | 0.004 | 0.000 |
| Indegree centralization | 2.8% | 0.4% | 0.8% | 0.7% | 0.1% | 1.3% | 2.9% | 6.4% | 11.1% | 0.3% | 0.9% | 0.2% |
| Outdegree centralization | 100.0% | 2.1% | 3.2% | 36.8% | 18.3% | 5.1% | 18.1% | 20.7% | 100.0% | 1.3% | 1.6% | 0.9% |
| Degree centralization | 50.0% | 1.1% | 1.6% | 15.5% | 9.4% | 2.5% | 9.3% | 10.3% | 50.0% | 0.7% | 0.7% | 0.5% |
| Betweenness centralization | 0.0% | 0.5% | 1.1% | 5.6% | 2.3% | 2.4% | 4.1% | 4.0% | 0.0% | 0.2% | 0.6% | 0.3% |
| Share of the main component | 100% | 100% | 100% | 100% | 100% | 100% | 100% | 100% | 100% | 96% | 98% | 95% |

**Table 10**: Structural characteristics for chain networks, link-tracing network and the network of networks. *Note:* [a] Measurements for these networks do not have a substantial meaning due to their small number of nodes. However, we did the computations for illustrative purposes. Readers should address the corresponding measurements with caution. [b] A variant of the link-tracing network wherein only the participants (the respondents) were kept. [c] The network level measurements should be interpreted in association with the number of basic elements in each network. [d] In the *Network of networks,* we filtered out the 2,540 symmetric ties (alter-alter ties).

## 4.6 Remarks on the representativeness of the sample data

There is mixed evidence regarding the representativeness of the sample estimates for the parameters in the populations of interests (Table 11). In the *Dâmbovița* subsample, the average age of participants was 37 (population mean: 48), while in *Castellón* it was 44 (three years more compared to the population mean). For gender, we found that 47% of the *Dâmbovița* subsample were female (four percent less compared to the population) whereas in *Castellón* we had 72% female participants (compared to 53% in population). However, it is noteworthy that the estimates are computed based on very small samples (154 respondents in *Dâmbovița* and 149 in *Castellón*). Taking into account the population size in each place, in a classic non-network probability survey, similar samples would return estimates within a confidence interval of eight (95% confidence level). From this perspective, some estimates delivered by the ORBITS link-tracing are very precise (for gender in *Dâmbovița* and for age in *Castellón*).

|  | Spain (Castellón) | Romania (Dâmbovița) |
|---|---|---|
| *Estimates and parameters\** | | |
| *Average age* | | |
| Population | 41 | 48 |
| Study participants | 44 (14) | 37 (17)\*\* |
| Seed 2 chain network | 45 (13) | 48 (12) |
| *Share of females* | | |
| Population | 53% | 51% |
| Study participants | 72%\*\* | 47% |
| Seed 2 chain network | 69%\*\* | 49% |
| *Size for populations of interest* | | |
| Population | 16,840 | 406,598 |
| Study participants | 149 | 154 |
| Seed 2 chain network | 107 | 77 |

**Table 11:** Estimates and parameters for age and sex in the populations of interests. *Note:* \* For individuals of at least 18 years old. \*\* Estimates with a deviation from the population mean higher than expected.

Given the lack of literature reporting results on the representativeness of empirical (and not simulated) link-tracing studies, we cannot but speculate on the possible factors responsible for estimate deviations. If sex homophily affects the estimates, then we should find better statistics in chain-networks that do not exhibit this clustering effect. Interestingly, the scores computed on the second seed chain network with smaller samples (this network does not display differential homophily, see Table 8) indicates good estimates. Classic surveys would return values within a 9.45 confidence interval (95% confidence level) for sample sizes similar to the one in Castellón, and within 11.2 (95% confidence level) for sample sizes similar to the one in Dâmbovița). However, the gender estimate in *Castellón* still severely deviates from the population mean. That is indicative of unobserved factors are at play. One of this possible factors might be the *masking effect* (in *Castellón,* 62% of the people elicited by the participants in their personal networks were not provided as a referral, *i.e.,* nominated as a possible participant to the research, compared to 39% in Dâmbovița). At the same time, in Dâmbovița, the deviation of the subsample from the population parameter might be the result of age homophily (39% of the people nominated as a referral were also elicited as a contact in the participants' personal networks).

# 5 Discussion and conclusions

In this paper, we present the results obtained after sampling from a hidden population, *i.e.,* people living in the transnational social field encompassing a Romanian migration destination place (*Castellón*, Spain) and their connections living in a Romanian migration sending community (*Dâmbovița*). We employed a link-tracing methodology built on a binational community approach. The collected data have a multi-layered structure. The first layer consists of 303 personal networks of migrants in Spain (147), returnees (18) and people with no Spanish migration experience (138). The second layer consists of a link-tracing network of 1,068 nodes and 1,187 referee-referral ties. Almost all nodes (96%) share membership in the same component. This network is the product of interconnecting the 303 respondents as well as other referrals who did not participate to the research. The third layer, the network of networks, was generated by integrating and interconnecting the participants, their referrals, the personal contacts (relatives, friends and acquaintances) and ties among these contacts. In the end, this network of networks consisted of 4,855 nominated people, 5,477 directed ties (nominations) and 2,540 undirected edges (alter-alter ties).

Our results indicate that differential homophily (on country and sex) significantly influenced nomination patterns (in the link-tracing network of participants), consistent with earlier research (Krivitsky et al. 2009; McPherson et al. 2001). For instance, in a similar study, Merli et al. (2016) report, for the case of Chinese migrants in Tanzania, that nominations were patterned by province of origin and ownership sector of employment.

Supplementary, separate chain-networks (resulted by decomposing the link-tracing network by seed) vary in residence and gender homophily. In the largest chain-networks, homophily is either absent, or present but under different forms. Exploring the structural characteristics of the nomination ties, we understood that, for instance, the network of networks and the link-tracing network display low levels of nomination reciprocity (less than .1%). The lack of reciprocity was detected to the other valid (chain) networks. Also, the number of ties observed in the link-tracing network is smaller compared to the one expected by chance alone. Additionally, migrant networks do not exhibit centralization, which means that brokers or central nodes are either non-existent or were not unveiled.

Concerning the representativeness of the link-tracing sample statistics, we report mixed evidence. On the one hand, given the small size of the samples, we provided good estimates for the parameters in the populations of interests without weighting (for average age and share of women). On the other hand, the estimation for gender in one of the two sites, Castellón, severely deviated from the mean. Empirical link-tracing studies on migration flows are rather nascent, so further research is needed to explore which factors impact on the estimates of the link-tracing samples.

The practical challenges of implementing binational link-tracing designs imply both technical (study-related) and contextual (external to the research) aspects. The selection procedures and the complexity of the data collection instruments assume a certain level of understanding (or education) and cooperation from the targeted respondents as well as their access to communicational technologies (mobile or smart phones for eliciting alters and nominations). Multi-sited research, by its nature, is conducted in different cultural and social contexts and under distinct institutional arrangements. Respondents' behavior and social realities radically change from one site to the other. Consequently, these require different field-work strategies from the part of the researchers, while operating under the same methodology. All these challenges have an effect on the response rate and the quality of the collected data.

With regard to the composition of the personal networks, data show that participants' social lives are dominated by family and friends in close proximity. Migrants have more ties in the destination, whereas returnees and non-migrants, in the origin. Furthermore, we provide evidence that social connectivity is not restricted to the binational corridor. Both Romanians living in Romania and living in Spain report a considerable number of connections to relatives and friends worldwide.

Our study falls within the emerging efforts of linking social network research to the investigation of international migration (Bilecen and Sienkiewicz 2015; Cachia and Maya Jariego 2018; Dahinden 2009; Hosnedlová 2017; Lubbers et al. 2010; Mazzucato 2009; Vacca et al. 2018). Specifically, the findings we report here on the Romanian transnational migration to Spain add to the cumulating data already produced, by similar research approaches, on Mexican-US cross-border migration (Mouw et al. 2014; Verdery et al. 2018) and Chinese migration to Tanzania (Merli et al. 2016). This might be beneficial both methodologically (*e.g.,* new insights on the robustness of link-tracing sampling from migrant populations) and substantially (*e.g.,* the possibility of comparing different international migration flows and of analyzing how migrants and non-migrants are embedded in transnational networks).

Based on our fieldwork experience, we consider that link-tracing sampling from migrant populations can be improved. To our knowledge, currently available software packages are not customized for the process of generating and controlling acronyms used for data anonymization. Human creation and manipulation of acronyms is prone to errors which, by consequence, it increases the necessary time for data cleaning and processing, and threatens the accuracy of datasets. This aspect is of critical importance especially for the creation of the transnational networks by inter-connecting the personal networks of the study participants. Due to the high specificity of sampling from TSFs (*e.g.,* multi-sited simultaneously data collection undertaken by international teams), existent software package tools, such as *RecordLinkage* in R, have only a limited efficacy in assessing and solving for conflicting acronyms (*e.g.,* cases wherein one individual has different acronyms, or several individuals have the same acronym).

We suggest several avenues for future research. Firstly, data collection tools should be further developed to provide a higher quantity of information while controlling for respondent burden. Secondly, the binational link-tracing methodology would need to be replicated and compared to other transnational migration flows. In this regard, we currently replicate the methodology to a different immigrant sending community in Romania (Bistrița-Năsăud) and destination area in Spain (Roquetas de Mar). Thirdly, research should assess the impact institutions and organizations have on supporting specific transnational networks of migrants (the multi-level organization of transnational networks) (Molina et al. 2018). Fourthly, the data collection process could be extended from a binational link-tracing sampling approach to a multi-national frame-work (conducting interviews in more than two countries). Fifth, collecting longitudinal data would increase the understanding of how *multi-level multinational networks* develop over time (multi-layered networks embedding different social entities – individuals and organizations, identified in multiple countries and connected by transnational relationships). And last, mathematical and statistical models should be customized to the specific nature of international migration.

## Acknowledgements

We gratefully acknowledge Iulian Oană, Bianca-Elena Mihăilă, Renáta Hosnedlová, Ignacio Fradejas-García, and Alexandra-Adelina Stoica, for their valuable contribution to this paper (data preparation and processing).

## Funding statement

This research was supported by the Ministry of Economy, Industry and Competitiveness – Government of Spain (MINECO-FEDER) (CSO2015-68687-P, 2016-2020) and by The Executive Agency for Higher Education, Research, Development and Innovation Funding (UEFISCDI – Government of Romania) (code: PN-III-P1-1.1-TE-2016-0362).